\documentclass[runningheads,a4paper]{llncs}
\usepackage{amssymb}
\usepackage{graphicx}
\usepackage{url}
 \usepackage{xspace}
\usepackage[resetlabels]{multibib}
 \usepackage{fancyhdr}
 \usepackage{morewrites}
\usepackage{enumitem}
 \usepackage{fancyhdr}

\newcites{farrar}{References}
\newcites{malviya}{References}
\newcites{vierhauser}{References}
\newcites{sharif}{References}
\newcites{guo}{References}
\newcites{chitre}{References}
\newcites{steghoefer}{References}
\newcites{hayes}{References}
\newcites{mahmout}{References}
\newcites{rahimi}{References}
\newcites{berger}{References}
\newcites{zoogan}{References}
\newcites{goodrun}{References}
\newcites{alenazi}{References}
\newcites{niu}{References}
\newcites{maeder}{References}
\newcites{goodrum}{References}
\newcites{lin}{References}
\newcites{liu}{References}
\newcites{zhao}{References}
\newcites{sessionFourIntro}{References}
\newcites{sessionThreeIntro}{References}
\newcites{sessionFiveIntro}{References}
\newcites{breakoutIndustry}{References}
\newcites{intro}{References}

\newcommand{\sessionheader}[1]{{\bfseries\fontfamily{pbk}\selectfont\noindent{{#1}}}}

\usepackage{etoolbox}

\newcommand*\maketitleses{%
    \begingroup
    \centering
    \fontsize{72}{80}
    \fontfamily{qhv}
    \fontseries{b}
    \fontshape{sl}
    \selectfont
    \headlinecolor
    \@title
    \par
    \vskip1in
    \endgroup
}

\renewcommand{\part}[1]{\addcontentsline{toc}{part}{#1}}

\newcommand{\context}{\hfill \break\textbf{[Context and Motivation]}\xspace\\}
\newcommand{\questions}{\hfill\break\textbf{[Question/problem]}\xspace\\}
\newcommand{\ideas}{\hfill\break\textbf{[Principal ideas/results]}\xspace\\}
\newcommand{\contrib}{\hfill\break\textbf{[Contribution]}\xspace\\}
\newcommand{\future}{\hfill\break\textbf{[Future Directions]}\xspace\\}
\newcommand{\probstate}{\hfill\break\textbf{[Problem Statement]}\xspace\\}
\newcommand{\contribfuture}{\hfill\break\textbf{[Contributions and Future Directions]}\xspace\\}

\newcommand{\method}{\hfill\break\textbf{[Method]}\xspace\\}
\newcommand{\results}{\hfill\break\textbf{[Results]}\xspace\\}

\newcommand{\ack}{\hfill\break\textbf{Acknowledgements}\xspace\\}

\usepackage{etoolbox}

\let\oldthesection=\thesection

\renewcommand{\thesection}{}

\makeatletter
\pretocmd{\chapter}{\addtocontents{toc}{\protect\addvspace{15\p@}}}{}{}
\pretocmd{\section}{\addtocontents{toc}{\protect\addvspace{5\p@}}}{}{}
\makeatother

\begin{document}

\mainmatter 
\title{\huge{Grand Challenges of Traceability: \\ The Next Ten Years}}
\titlerunning{Grand Challenges of Traceability 2017}

\author{Giuliano Antoniol \and Jane Cleland-Huang  \and \\ Jane Huffman Hayes\and Michael Vierhauser}
\authorrunning{Grand Challenges of Traceability 2017}

\institute{\phantom{xxx}}





\let\oldaddcontentsline\addcontentsline
\def\addcontentsline#1#2#3{}
\maketitle
\def\addcontentsline#1#2#3{\oldaddcontentsline{#1}{#2}{#3}}

\includegraphics[width=0.8\textwidth]{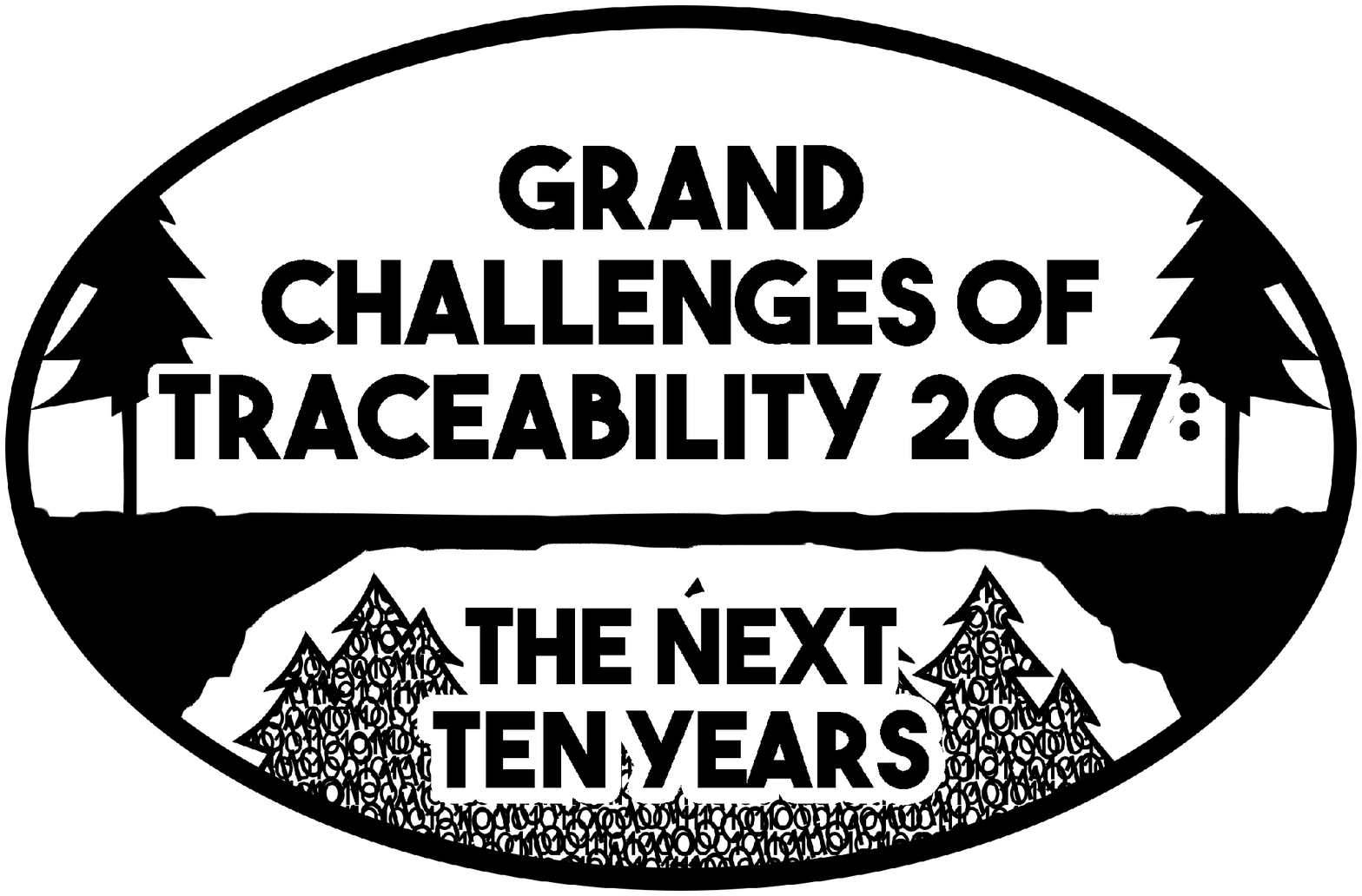}

\tableofcontents

\newpage

\part{The Grand Challenges Revisited - Then and Now}

{\fontfamily{qtm}\fontsize{22}{15}\selectfont 
{\noindent\Huge{Organizing Committee}}
}

\bigskip
\bigskip
\normalsize{}
\noindent \textbf{Alexander Dekhtyar}\\
Department of Computer Science \\
California Polytechnic State University\\
\\
\noindent \textbf{Bonita Sharif} \\
Department of Computer Science and Information Systems\\ Youngstown State University\\
\\
\noindent \textbf{Giuliano (Giulio)  Antoniol}\\
\emph{Program Co-Chair}\\
Ecole Polytechnique de Montreal\\
\\
\noindent \textbf{Jane Cleland-Huang}\\
\emph{Program Co-Chair}\\
Department of Computer Science and Engineering\\
University of Notre Dame\\
\\
\noindent \textbf{Jane Hayes}\\
\emph{General Chair}\\
Department of Computer Science\\
University of Kentucky\\
\\
\noindent \textbf{Michael Vierhauser}\\
\emph{Proceedings Chair}\\
Department of Computer Science and Engineering\\
University of Notre Dame\\
\\
\noindent \textbf{Nan Niu}\\
Department of EECS\\
University of Cincinnati\\
\\
\noindent \textbf{Tingting Yu}\\
\emph{Local Organizer}\\
Department of Computer Science\\
University of Kentucky\\
\bigskip



\newpage

{\fontfamily{qtm}\fontsize{22}{15}\selectfont 
{\noindent{Grand Challenges of Traceability 2017}}
}
\\
\\
\noindent Giulio Antoniol, Jane Cleland-Huang, Jane Huffman Hayes, Michael Vierhauser\\

\author{Giulio Antoniol, Jane Cleland-Huang, Jane Huffman Hayes, Michael Vierhauser}

{\section{The Grand Challenges Revisited}}
Just over ten years ago a group of researchers met under the St. Louis Arches at the 2005 International Conference on Software Engineering and formulated a plan to launch the Center of Excellence for Software Traceability.  The goal was rather audacious -- to identify the Grand Challenges of Traceability and to work together to forge real, industrial strength solutions that would advance the field in non-trivial ways.

Stepping back a few years to the early 1990s, Olly Gotel and Anthony Finkelstein had conducted an extensive study with industrial practitioners and published what has now become a seminal paper in the field, entitled ``An Analysis of the Requirements Traceability Problem''
\citeintro{Gotel-Finkelstein}. Their work highlighted several traceability challenges -- especially those related to the practice and processes of establishing traceability in an industrial setting. In another seminal paper published in 2001, Ramesh Balasubramaniam conducted an indepth study of requirements traceability in practice and constructed traceability metamodels for various software engineering tasks \citeintro{Ramesh2001MetaModel}. These papers, alongside others that emerged in the early 2000s, clearly defined the traceability problem and laid a solid foundation for a flurry of research that has continued until this day.

The year 2002 also marked a significant step forward in the landscape of traceability research when Giulio Antoniol, Gerardo Canfora, Gerardo Casazza, Andrea De Lucia, and Ettore Merlo published a paper describing the use of information retrieval techniques to automatically construct trace links between documentation and code \citeintro{Antoniol:Recovering}. This paper, in particular, attracted numerous researchers to the field with the vision of full automation.

This led us to 2005 and the St. Louis arches -- when Jonathan Maletic, Giulio Antoniol, Alex Dehktyar, Jane Huffman Hayes, and Jane Cleland-Huang conceived of the idea of forming the Center of Excellence for Software Traceability.  The idea was to harness the energy and enthusiasm of the research community, and to work together to evaluate the state of practice and identify open challenges to be addressed. This ultimately led to a series of workshops, the first of which was hosted at NASA's IV\&V facility in Morgantown, West Virginia in 2006, and the second at the Natural Bridge State Park outside Lexington, Kentucky in 2007. The meetings resulted in an initial document outlining open traceability problems.

As a result of these meetings, a group of researchers worked together over a period of several years to formulate the Grand Challenges of Traceability \citeintro{DBLP:books/daglib/p/GotelCHZEGDAM12} and a subsequent traceability road map paper entitled ``The quest for ubiquity: A roadmap for software and systems traceability research'' \citeintro{GotelCHZEGA12}. The Grand Challenges centered around the quality goals of traceability -- and included the need for traceability to be purposed, cost-effective, configurable, trusted, scalable, portable, valued, and ultimately ubiquitous.  This final goal was defined as the Grand Challenge of Traceability and stated that ``Traceability is always there, without ever having to think about getting it there, as it is built into the engineering process: traceability has effectively 'disappeared without a trace'.''
The goal-orientation of the original Grand Challenges was inspirational, but failed to fully describe the practical and technical research challenges that our community needed to address in order to achieve our goal of ubiquitous traceability.  

Therefore, a smaller group of researchers published another paper entitled ``Software Traceability: Trends and Future Directions'' \citeintro{DBLP:conf/icse/2014fose} in the 2014 Future of Software Engineering track at the International Conference on Software Engineering.  This paper identified specific research areas and mapped them to quality goals, traceability processes, and open technical challenges.  These challenges were grouped into three main areas of planning and managing, creating and maintaining traces, and using traces.  Specific research directions focused on understanding stakeholder needs, strategizing, trace link creation, trace maintenance, trace integrity, querying across trace data, and visualization, all of which represent active areas of research today.  The abstracts presented in this proceedings cover many of these areas and provide insights into several current research projects. 

In our first Natural Bridge symposium, held in 2007, our research community enthusiastically set out to address the Grand Challenges of Traceability.  Now, ten years later, we came together again to ask some hard questions:  How are we doing?  Are we getting there?  What still needs to be done? These proceedings attempt to answer those questions. They include a series of short position papers, representing some of the current work in our community organized across the four process axes of traceability practice \citeintro{Gotel2012}. 

Researchers have developed tools and techniques to help users plan and strategize their tracing solutions. Some of these solutions were presented in Session 1 on \emph{Trace Strategizing} led by Patrick M\"ader and Nan Niu. Research in automating the creation and maintenance of trace links has been particularly active. Current projects and results were reported in Session 2 on \emph{Trace Link Creation and Evolution} organized by  Giuliano Antoniol and Jin Guo. In the area of \emph{Trace Link Usage} there have been a number of critical, in-depth studies over the past decade, which have provided important insights into the traceability needs and practices of users. These were discussed in Session 3, coordinated by Markus Borg and Bonita Sharif. Session 4, led by Sahar Kokaly and Michael Vierhauser, discussed \emph{real-world applications of Traceability}.  Finally, in Session 5, Mona Rahimi and Carlos Bernal-C\'ardenas led a panel discussion on \emph{Traceability Datasets and benchmarks}.

Through interactive discussions, participants at the 2017 Grand Challenges of Traceability event identified two specific challenges that must be addressed as we move forward into the next ten years of research. The first relates to the availability of traceability datasets and benchmarks, while the second relates to real-world applications of traceability. Ten years ago, when some of us met under the St. Louis arches, the community had access to two datasets, collected and shared by Jane Huffman Hayes.  While many of our community benefited from these datasets, they were rather small,  and hardly led to experiments with generalizable results. Today, we have over 16 datasets available for community use from our CoEST.org website; however, as discussed in the section on \emph{Discussion and Breakout Groups} we still have a long way to go.  Participants also discussed challenges related to the adoption of tracing techniques in industrial practice. Industrial and governmental organizations continue to struggle with tasks that could be alleviated by more ubiquitous use of traceability -- but tools and techniques from the research community have not effectively infiltrated industrial practice.  Issues and potential solutions are described in  the section on \emph{Industry Transfer}.

Looking forward, we are encouraged by the many active, ongoing, and impactful research projects that members of the traceability community are engaging in.  Our hope is that ten years from now we will be able to look back at a productive decade of research and claim that traceability is always present, built into the engineering process, and has ``effectively disappeared without a trace'' \citeintro{DBLP:conf/re/GotelCHZEGA12}.  We hope that others will see the potential that traceability has for empowering software and systems engineers to develop higher-quality products at increasing levels of complexity and scale, and that they will join the active community of Software and Systems traceability researchers as we move forward into the next decade of research.

\phantom{xxx}

\phantom{xxx}

\phantom{xxx}

\bibliographystyleintro{modifiedplain}
\bibliographyintro{abstracts/bibs/intro}

\newpage
\part{Trace Strategizing}
\author{Session Chairs:\\Patrick M{\"{a}}der and Nan Niu}
\title{\sessionheader{Session 1:\\ Trace Strategizing}}
\authorrunning{Session 1}
\label{sec:Strategizing}
\institute{}
\maketitle
\setcounter{section}{0}
\vspace{40pt}

\label{sess:strate}

This session focuses on \emph{trace strategizing}, i.e., the plan and management of a traceability strategy which underpins and cuts across the creation, use, and maintenance of the trace links. Not only does a traceability strategy have to be planned at the beginning of a project to instrument the tracing environment, but this strategy also evolves as the project proceeds to better guide the various tracing activities and meet the stakeholder goals.

In this session, we began with Patrick's presentation where he introduced the importance of trace strategizing and shared his research experience in related topics such as traceability information model (TIM) used in safety-critical projects, costs and benefits of traceability, and so on. His chair's message was kept short, aiming to set the stage for the focused talks and to stimulate broader participation.

Three abstracts were grouped in this session, and the authors presented their work after Patrick's talk. We asked the presenters beforehand to prepare a few open-ended and even controversial questions derived from their work that they would like to engage the workshop participants. We then clustered the discussion topics based on the presenters' preparations and the notes taken during the presentations, asked each workshop participant to join a discussion group where they could exchange their views and visions, and finally, shared each group's discussion highlights to the entire workshop.

We summarize the discussion highlights from each of the four groups as follows.

\begin{itemize}
    \item Evaluation of traceability \\
    
    The group discussed the utility aspects of assessing traceability methods, techniques, and tools. While metrics like recall and precision were important, the group suggested the idea of assessing traceability utility in specific activities and tasks, e.g., speed-up. A specific point was raised in terms of repeated studies (or a meta study) where the users would be asked to tell what they think the traceability results are good to them. \\

    \item Traceability information model (TIM) \\
    
    The definition and evolution of a TIM were discussed among the group members. They talked about defining the TIM in a formal way versus a less formal way, probably depending on the application domains and development processes (e.g., safety-critical systems versus agile projects). The group also raised the question of how to (best) define the TIM: should it be based on artifacts, on tasks, and/or on goals? \\
    
    \item After-the-fact tracing \\
    
    The group reported some challenges of doing after-the-fact tracing, including the handling of missing links would cost much effort of identifying and fixing them. Meanwhile, advantages like taking one or two people off the project team (as opposed to finding random persons) to perform after-the-fact tracing were also discussed. \\

    \item Better humans \\
    
    The focal point of the group was how to educate and train the next generation of software engineers so that they could be better at traceability. Specific teaching practices were exchanged, such as applying penalty if the students' code does not trace back to the features of their own choices. Much debate was around the value and the cost of traceability, e.g., the return on investment (ROI) of tools like GitHub and JIRA was short-term (committing the code change to close an issue returns an immediate value), whereas the ROI of traceability could be rather long-term (certification after a project milestone is reached). \\

\end{itemize}

As session chairs, we thoroughly enjoyed the presentations and the discussions on trace strategizing, and would like to specifically thank Jane \& Jane for giving us the opportunity to lead the kick off of the very interactive and fruitful two-day events at GCT-10. Our own notes from the session recorded more questions than answers regarding the traceability strategy: how to create one (generalizability versus customizability)? how to evaluate it (e.g., completeness)? what happens when it evolves? what should be the cost-benefit considerations when creating, evolving, and using it? \ldots The list goes on. We hope to see more work done in the next ten years to tackle the many facets of trace strategizing.



\newpage
\title{Automated Requirements Traceability}
\subtitle{The Study of Human Experts}

\author{Bhushan Chitre\inst{1}\and Jane Huffman Hayes\inst{1} \and Alex Dekhtyar\inst{2} \and Vivian Fong\inst{2}}
\authorrunning{Chitre et al.}

\institute{Computer Science Department\\
	University of Kentucky,\\
	Lexington, Kentucky, USA\\
	\email{bhushan.chitre@uky.edu, hayes@cs.uky.edu}
	\and
	Computer Science Department\\
	California Polytechnic State University,\\
	San Luis Obispo, CA, USA\\
	\email{\{dekhtyar, vfong01\}@calpoly.edu}
}

\maketitle

\context 
Experimentation in software engineering in general, and in traceability specifically, is hard \citechitre{7}. It requires a community of researchers that can replicate studies, experiments, abstract models to verify the results and extract observations on the discipline. We know that humans are fallible from the past experiments. In previous experiment \citechitre{1} our assumption was that the quality of the starting TM would impact the final TM and we thought the human analysts would make the final TM better but the study showed that to be false. Now in this paper, we present a replication of a traceability experiment with a slight modification to the original experiment \citechitre{1}. We apply the techniques from the previous experiment yet we have implemented them as TraceLab (TL) components \citechitre{1}. Research has shown that information retrieval techniques can be effectively applied to generate a candidate traceability matrix (TM) in an automated fashion for textual artifacts \citechitre{1,2,3}. Automated methods generate TMs that must be examined by human analysts - they must add and remove links as necessary to arrive at the final TM. \citechitre{5} In the prior experiment, each participant was given a download link to the tracing tool with the experiment components, library packages, instructions on how to use and install it and also a training dataset to get them familiar with the tool. Then they worked on their own time to trace the experimental dataset. Once they finished the given task, they sent their time to perform the task and final TM via email to the research team. We plan to run the experiment in this same manner in computer science upper and lower division software engineering classes at the University of Kentucky and California Polytechnic State University. 

\probstate
Our main goal of the experiment is to find out if human analysts must correct automatically generated trace matrices. We question if the quality of the candidate TM and time spent by the analyst influences the final TM quality.

\ideas The main goals are: human study with candidate TMs of varying quality, comparing the results to the original experiment, and taking the stand-alone tool and re-implementing it in TraceLab. The previous experiment used RETRO.NET \citechitre{4,6} and we plan to implement that as TraceLab components for our experiment and compare the results. Also, we want to develop a lab package or some guidance/suggestions for contributors to the previous experiment on lessons learned while implementing the tracing experiment in Tracelab, suggestions to future Tracelab developers who wish to work on TL components and other experiments.

\contribfuture
We want to improve better understanding of learning about the accuracy of humans with candidate TMs. And also, we plan to expand the development of converting stand-alone tools into TraceLab experiment components.

\bibliographystylechitre{modifiedplain}
\bibliographychitre{abstracts/bibs/chitre}

\newpage
\title{Best of Both Worlds: Synthesizing the Human and Method Sides of Requirements Tracing}

\author{Nan Niu\inst{1} \and Juha Savolainen\inst{2} \and Wentao Wang\inst{1} \and Mounifah Alenazi\inst{1}}

\institute{Department of Electrical Engineering and Computing Systems,\\
        University of Cincinnati, USA\\
	\email{nan.niu@uc.edu, wang2wt@mail.uc.edu, alenazmh@mail.uc.edu}
	\and
IGlobal Software and Control R\&D,\\
Danfoss Drives A/S, Denmark\\
	\email{ juhaerik.savolainen@danfoss.com}
}

\authorrunning{Niu et al.}
\titlerunning{Best of Both Worlds}

\maketitle

\context
In requirements tracing, the study of methods refers to the ``after the fact''\footnote{This assumption assumes that the two artifacts under tracing represent the final versions and do not change over time~\citeniu{Hayes-TSE06}.} recovery of candidate traceability links via automated methods---most notably the algorithms developed in information retrieval (IR)~\citeniu{Hayes-TSE06}. Not only do the different IR-based methods have a comparable trace retrieval performance, but the assumption of ``after the fact'' also limits the \emph{use} of traceability to assisting in tasks like change impact analysis rather than supporting the actual change (e.g., applying source code edits, checking architectural conformance, suggesting refactoring, etc.).

The seminal work by Hayes and Dekhtyar~\citeniu{Hayes-TEFSE05} triggered a series of studies of the human side of requirements tracing, including statistically testing the variables that might affect analyst performance~\citeniu{Dekhtyar-RE11}, analytically defining the measures to capture analyst work progress~\citeniu{Kong-RE12}, and theoretically understanding the mechanisms underlying analyst behavior~\citeniu{Niu-ICSE13}.

\questions
To truly gain the best of both worlds that will allow human and automated tool to do what they do best~\citeniu{Dekhtyar-RE11}, focusing only on either side is insufficient. As far as the method side is concerned, for example, assuming one and only one ``answer set'' to evaluate the trace retrieval effectiveness is recently challenged~\citeniu{Murugesan-RE16,Niu-FSE16}. As for the human side, we should not observe only human's reactions to a fixed tracing tool, but study their interactions with a wide variety of tool design factors, ranging from preprocessing steps and parameter setting to trace link visualization and use \emph{in situ}.

\ideas
The idea of \emph{complementarity} is appealing to us in that it leads to the synergistic effect where we can have ``the whole is greater than the sum of its parts''---namely, achieving the best of both worlds in requirements tracing. Our recent \emph{a posteriori} analysis of human analysts' tracing logs exploited associations and also considered contextual variables which the complementarity might be sensitive to.

\contrib
Our main contribution so far is the realization that complementarity needs to be rigorously tested. Our log analysis mentioned above involved four steps: (1) defining the objective function that quantifies the marginal return of the complementary activities, (2) removing insignificant activity combinations via association support, (3) classifying complementary relations based on association confidence, and (4) detecting contextual variables by regression tests. From all the possible 55 pairs of requirements tracing practices that we investigated, the analysis following the above 4 steps accepted only 6 of them, showing that achieving the best of both worlds is a nontrivial matter but a highly selective process.

\future
We echo Cleland-Huang \emph{et al.}~\citeniu{JCH-FOSE14-Short} that one ongoing research thrust is on making the traceability information accessible to humans to support the tasks that are relevant to their project environments, and making it rendered in ways that facilitate interaction and decision-making. Incorporating complementarity into the future vision requires the traceability benchmarks include not only artifacts' information (e.g., what counts as a link and what does not) but also fine-grained human interaction data (e.g., tasks performed, candidate links probed, learning materials used, solutions, rationales, etc.). Only by building and sharing rich, human-centric traceability datasets can we turn the best of both worlds from an \emph{a posteriori} analysis to a wisdom \emph{a priori}.

\bibliographystyleniu{modifiedplain}
\bibliographyniu{abstracts/bibs/niu}

\newpage

\title{Traceability Queries and Strategies for the Requirements Engineering Domain}
\author{Sugandha Malviya\inst{1} \and Michael Vierhauser\inst{2} \and Jane Cleland-Huang\inst{2}}
\authorrunning{Malviya et al.}
\titlerunning{Traceability Queries for the RE Domain}

\institute{School of Computing\\DePaul University,  Chicago,  USA.\\
	\email{slohar@depaul.edu}
	\and
	Department of Computer Science and Engineering\\
	University of Notre Dame, South Bend, IN, USA\\
	\email{mvierhauser@nd.edu, janeclelandhuang@nd.edu}
}

\maketitle

\context Requirements Engineering (RE)  is a vital process for creating high quality software systems~\citemalviya{hull2010requirements,hofmann2001requirements} comprising diverse tasks related to discovering, documenting,  and maintaining different kinds of requirements.  A plethora of different tools,  methods,  and techniques  are needed to successfully perform these tasks; however, even with proper tool-support in place,   they can be time-consuming and difficult to perform. Researchers have identified numerous problems associated with performing tasks as diverse as stakeholder identification, requirements elicitation and analysis, requirements specification and change management.

\questions One of the major obstacles in supporting such RE techniques stems from the fact that information needs to be collected and consolidated from many different and diverse data sources. For example,  performing {\emph stakeholder analysis} requires retrieving, collating, and analyzing information from interview notes,  scenarios,  goals and other artifacts~\citemalviya{zowghi2005requirements}. Typically, such artifacts are neither managed by a single person, nor stored in a single location but distributed across multiple repositories (e.g.,  document management systems,  source-code repositories,  issue trackers,  or application life-cycle management (ALM) services). Furthermore,  artifact data is sometimes incomplete,  inconsistent and important trace links can be missing \citemalviya{Rempel14,Ramesh2001}.  Accessing these data sources and combining them to produce meaningful and desired results can therefore be considered a  cumbersome and error-prone  endeavor.  Such problems can best be alleviated by strategic upfront project planning and with appropriate instrumenting of the environment.
 
\ideas The first and foremost objective of our research is to uncover the information needs of a Business Analysts and/or Requirements Engineer for supporting different RE tasks. Several user studies ~\citemalviya{ko2007information,fritz2010using} have investigated and collected important questions asked by software practitioners in various project roles, and then analyzed the questions to discover the information snippets needed to perform their tasks. Analyzing real-world trace queries can shed light on the questions requirements professionals would like to ask and the artifacts needed to support such questions. In our recent work, we followed an empirical approach and identified 159 traceability queries by 29 requirements professionals in IT industry. Using open coding and grounded theory, we analyzed and grouped these queries into 9 different query purposes and 53 sub-purposes, and also identified frequently used artifacts across different query-purpose. The subsequent aim is to synthesize and showcase the gathered information by developing traceability models pertaining to different query goals. This is potentially useful for project-level planners, and could help them to identify important questions, proactively instrument their environments with supporting tools, and strategically collect data that is needed to answer the queries of interest to their project.  

\contrib
As part of our work we collected and constructed a traceability query-set representing the questions requirements professionals are interested in asking. Our ongoing work furthermore provides the foundation for creating traceability information models(TIM)\citemalviya{FDASoftware} representing multiple aspects of requirements tasks. This knowledge can be used to plan traceability strategies in advance, which will support decision-making, project planning and in conducting the overall requirements activities. The final contribution of our research work is to provide an extensive information set that represents the RE domain.  

\future In our future work we plan to extend the collected set of queries including additional data sources and to thoroughly validate the generated TIM by conducting a usability study with requirements professionals.

\bibliographystylemalviya{modifiedplain}
\bibliographymalviya{abstracts/bibs/malviya}



\newpage

\part{Trace Link Creation and Evolution}
\author{Session Chairs:\\Giuliano Antoniol and Jin Guo}
\title{\sessionheader{Session 2:\\ Trace Link Creation and Evolution}}
\institute{}
\authorrunning{Session 2}
\titlerunning{Session 2}
\maketitle

\setcounter{section}{0}
\vspace{40pt}



In the past decades, researchers have devoted significant effort in the area of trace link creation and evolution. In order to automatically acquire high-quality trace links, numerous techniques have been proposed.  In this session, we gathered six contributions discussing the latest recovery and evolution techniques, contrasting and comparing pros and cons as well as open issues that need to be addressed in order to bring real breakthrough in this area for achieving ubiquitous traceability.

Despite recent visionary works, for trace link creation, the text analysis based methods are still the main stream.  These methods aim to construct trace links from analyzing the textual content of each individual software artifacts.  Many variants and modifications have been proposed to improve the conventional term-based trace retrieval approaches. Most promisingly, a new generation of deep-learning inspired methods starts to attract increasing attentions due to their noteworthy success in the natural language processing domain. In two abstracts, Zhao et al. and Guo et al. both proposed utilizing deep learning techniques to improve the trace link retrieval results for software artifacts. However, unlike other domains in which big data are relatively easy to collect, software traceability problems are normally plagued with a limited amount of available data.  This fact brings to the key question: will those trendy approaches once adapted to traceability problems return generate trace links with satisfactory precision and recall? Results by Guo et al. are very encouraging and point in that direction. Still, in their abstract, Hayes et al. raised several questions related to adapting and applying deep learning to tracing problems for our community to discuss. 

Other than the textual content of software artifact, information from other sources can be extremely effective to create and maintain trace links. For example, Rahimi et al. utilized source code change patterns to evolve trace links across different software versions.  Sharif proposed deploying eye tracker to collect and analyze developers' eye gazes when they are at work. The eye tracking data, in turn, can be used to create and maintain trace links during the software development process with no extra or minimal effort. 

Finally, as discussed in the abstract by Alenazi et al., we are currently observing a deeper relation between software and mechatronic components, parts or systems. This raised the challenging problem of properly identifying the tracing target.  Indeed, this is non trivial but especially critical for systems that require seamlessly interactions between software and mechatronic parts. 

We hope that you will enjoy the contributions from this session. We see the diverse methods and important issues discussed in this session as stimulation for our whole community to explore the future directions for meeting the grand traceability challenge from the perspectives of trace link creation and evolution.

\vspace{0.5cm}


Heated discussions around different aspects of trace link creation and evolution techniques took place after the brief presentations of each abstract. Participants engaged in passionate discussion, focusing on various challenges. Special attention was given to the following topics:

\begin{itemize}[leftmargin=0cm,itemindent=.5cm,labelwidth=\itemindent,labelsep=0cm,align=left]
\item Should we put faith in deep learning? How to solve their dependencies on large amount of training data? How to improve the generalizability of automated tracing solutions?\\
Around these issues, participants demonstrated their great interest despite the fact that they also raised some reasonable doubts. For example, on the one hand, the pre-trained deep learning models provide opportunities of utilizing data from the general domain in different software engineering application domains. This advance would potentially increase the generalizability of the trace retrieval solutions. However, it is not clear if the generic models will need adaptation or to what extent they will generalize to specific domains. On the other hand, all participants agreed that our communities need to make great effort towards building benchmarks and sharing the data and models to enable replicating and advancing the proposed tracing solutions in various situations. 
\vspace{3mm}
\item How to establishing proper datasets to evaluate trace link evolution techniques? What are the types for such dataset and how to decide on the granularity of trace links? Should the trace links expend their boundary to the mechatronics in which the software system is engineered?\\
Participants underlined that the dependency on dataset is not unique for deep–learning based algorithms. It is the foundation for any traceability problems. For trace link evolution specifically, the scope of tracing need to be clearly defined.  Some evolution datasets are created in class settings in universities.  But more effort is needed to acquire and share datasets aiming to better address the real world scenarios of software evolution. Concerns were also raised on a trend of limiting the research to open source software. Indeed, on one hand this simplifies data exchange. But on the other hand it may not be representative for several industrial sectors (e.g., aerospace, medical systems or automotive).
\vspace{3mm}
\item How important and realistic to capture the environmental data generated by human for the traceability purpose, such as process exhaust and eye gazes? \\
This information would undoubtedly generate great value for creating and maintaining trace links with during the software development process. But there are still open questions.  For example, how to persuade software developers to adopt new tools or processes, and is it feasible to ask developers to use an eye tracker during their daily work. Another point-of-concern is the general lacking of evidence of trace usefulness.  As a system ages, traces would become outdated. An eye tracker would make it easier to evolve traces. But it is still unclear how useful the evolved traces will be.  To find answers to these questions, more intensive studies of software developers and their working habits are essential as future research directions.
\end{itemize}

To summarize, we are still far from the utopia of creating and maintaining perfect trace links.  But the techniques and issues discussed in this session demonstrated potential directions for the next decades of traceability research. We hope you find this session fruitful and inspiring.



\newpage

\title{Using Deep Learning to Improve the Accuracy of 
Requirements to Code Traceability}

\authorrunning{Zhao et al.}
\titlerunning{Deep Learning for Req. to Code Traceability}

\author{Yu Zhao \and Tarannum S. Zaman \and Tingting Yu \and Jane Huffman Hayes}

\institute{Department of Computer Science\\
	University of Kentucky\\
	Lexington, Kentucky, 40506, USA\\
	\email{yzh355@g.uky.edu | tarannum.zaman@uky.edu | tyu@cs.uky.edu | hayes@cs.uky.edu}
}

\maketitle

\context
Information retrieval (IR) techniques have been used
to recover traceability links between natural language requirements
and source code. However, IR techniques are often lack of accuracy. 
To address this problem, research has shown that mining software repositories
and using the mined results combined with the IR techniques 
can improve the accuracy~\citezhao{Ali13,dit2013dataset}.  For example,
Histrace~\citezhao{Ali13} identifies 
traceability links between requirements and source code
through CVS/SVN change logs using a Vector Space Model (VSM). 
The log messages are tied to changed entities and, 
thus, can be used to infer traceability links.

\questions
While these approaches
are promising, they rely on the assumption that different
types of knowledge (e.g., commit messages, code comments)
of the repositories exist.  In many cases, however, such knowledge
may not be available. 
For example, 
code commenting has been a standard practice in software
development. Despite the need and importance of code comments, 
many code bases do not contain adequate comments~\citezhao{de2005study}. 
Another type of knowledge involves commit messages,
which have been used to document changes of software
in version control systems. However,
research~\citezhao{dyer2013boa}  have shown that 14\% of the commit
messages are empty and 66\% of the messages contain fewer
words than a typical English sentence. 
To address the above problems on inadequate documentation, research
 on automated natural language
text generation in software repositories have been
proposed. For example, 
Wong et al.~\citezhao{wong2013autocomment} generate comments
automatically by mining Question and Answer (Q\&A) for
code-comment mappings. However, this approach has several drawbacks. 
First, it cannot handle cases in which  
the text descriptions do not exist in the mapping database. 
Second,  there is not a notion of semantic similarity between 
words when generating the comments. 
Third, this approach is not scalable in the presence of large amount
of data involving the code-comment mappings.
Regarding the commit message generation,
ChangeScribe~\citezhao{cortes2014automatically} generates
commit messages  by taking into account the change types, such as
file rename and deletion. 
However, this message generation approach is based on  
pre-defined templates and thus may not represent
the real meanings of the changes. 

\ideas
In this research, we propose an approach to 
automatically generate natural language texts 
that can build the bridge to recover traceability links
between requirements and code.  We focus 
on commit message and code comments generation. 
To address the aforementioned challenges imposed by existing  
techniques, we employ the deep neural network (also 
known as deep learning), featured by its ability of learning highly
complicated features automatically~\citezhao{lecun2015deep}.
We propose to leverage  recurrent neural networks (RNNs), which are 
suitable for modeling texts (i.e., a sequence of characters) by its iterative nature. 
Natural language generation using RNNs differ from
text mining and retrieval systems; the generated descriptions
are different from any existing commit messages
or comments, which are more flexible and may accurately 
reflect the semantic meanings. 
We will use Web Crawler to craw HTMLs in the 
Question and Answer  (e.g., StackOverflow) and tutorial web sites (e.g., W3C). 
We can then utilize the natural language processing method to obtain the 
mapping between code and its corresponding descriptions. 
Next, we will train the RNNs by using these mappings. Specifically,
The source code is the input to the RNNs and the text description
(i.e., commit messages or comments) are the labels. 
Since today's software artifacts have become ``big data", 
the training data is sufficient. As such, 
it is possible to train a generative text model based on the source code.
Finally, given a code segment, the trained model  can generate
the corresponding text descriptions. 

\future
In this research, we propose to train deep neural networks  for 
generating text-based knowledge in software repositories to improve the
accuracy of traceability links recovery.
We will perform an  empirical study to evaluate 
our proposed approach. 
We envision several scenarios where deep neural networks 
may address long-standing  software engineering research challenges,
including automated program generation from natural
languages and test oracle generation.

\bibliographystylezhao{modifiedplain}
\bibliographyzhao{abstracts/bibs/zhao}

\newpage

\title{Too Little for Big Data? }

\author{Jane Huffman Hayes\inst{1} \and Giulio Antoniol\inst{2} \and Licong Cui\inst{1} \and Tingting Yu\inst{1}}
\authorrunning{Hayes et al.}

\institute{Computer Science Department\\
	University of Kentucky\\
	Lexington, Kentucky, USA\\
	\email{\{hayes, tyu\}@cs.uky.edu,\\  licong@uky.edu}
		\and
\'{E}cole Polytechnique de Montr\'{e}al\\
	Montreal, Quebec, Canada\\
	\email{antoniol@ieee.org@nd.edu}	
}

\maketitle

 \context
 Trace matrices are the lynchpin of verification and validation activities that must be performed for mission- and safety-critical software systems:  criticality analysis, completeness analysis, change impact analysis, etc.  Studies have shown that automated traceability techniques can achieve high recall and sometimes acceptable precision when used to generate trace matrices \citehayes{h1}. The human analyst is required in the loop for many critical software systems and plays a role in vetting the auto-generated trace matrices.  Studies have shown that humans are fallible and tend to decrease the accuracy of auto-generated trace matrices \citehayes{h2,h3,h4}.  To address the need for improved matrix quality and synergy with analysts, researchers are examining methods that have received popular and high acclaim.  We surmise that “big data,” deep learning, and meta-heuristic search are three categories of interest.   Big data refers to “an emerging data science paradigm of multi-dimensional information mining for scientific discovery and business analytics over large-scale infrastructure” \citehayes{h5}. In addition, when facing complex classification problems, deep learning \citehayes{h7} has proven to be effective \citehayes{h7,h8}. However, not all data have been created equal, and some data are likely more important than others \citehayes{h11}; unfortunately, exhaustive search is oftentimes not feasible and we must resort to heuristic methods \citehayes{h10}.  
 
\probstate
 Automated trace link generation techniques suffer from low precision and lack of synergy with human analysts.  There is a potential that big data technologies, deep learning, and heuristic optimization can assist with automated trace link generation due to enormous software artifacts data and its complex structures.
 
 \ideas
 We plan to characterize trace generation in terms of an unbalanced big data classification problem.  For example, we will examine the typical size of software engineering artifacts, software elements that comprise the artifacts, diversity of the datasets, granularity of the datasets, and align them with big data technique pre-requisites/requirements.  Though it may appear that traceability datasets are not large enough to apply big data techniques, with software engineering artifacts generally consisting of thousands of elements versus millions or billions, we can borrow semantically reach words encoding from natural language processing techniques \citehayes{h8}. Possibilities for addressing this include increased granularity in order to expand the size of datasets, deriving more data elements featuring disparate aspects of the datasets, etc. However, new ideas are needed to properly handle the challenge of highly unbalanced datasets where only a handful of true links exist.   We expect that expanding the size of the data could simplify the rebalancing problem and improve the accuracy of trace generation. A second possibility would be to formulate the classifier-rebalancing problem as a search problem \citehayes{h10} or to model the trace recovery as a classification task where deep learning techniques place true traces close in the feature space making the similarity between true links higher. Alternatively, we may use big data representations for trace elements such as directed acyclic graphs and perform concept mining over the graphs \citehayes{h6}. 
 
\contribfuture
 Inspired by prior work \citehayes{h10,h11,h12}, we plan to capitalize on big data approaches successfully applied to biomedical problems \citehayes{h6}, heuristic optimization \citehayes{h10}, and deep learning \citehayes{h7,h8,h12} and learn how to apply them to the trace link generation problem.

\ack
This work was supported by NSF grants CCF-1464032 and CCF-1511117.

\bibliographystylehayes{modifiedplain}
\bibliographyhayes{abstracts/bibs/hayes}

\newpage
\title{Traceability for Evolving \\Automated Production Systems}

\author{Mounifah Alenazi\inst{1} \and Nan Niu\inst{1} \and Wentao Wang\inst{1} \and Birgit Vogel-Heuser\inst{2}}

\institute{Department of Electrical Engineering and Computing Systems,\\
        University of Cincinnati, USA\\
	\email{alenazmh@mail.uc.edu, nan.niu@uc.edu, wang2wt@mail.uc.edu}
	\and
Institute of Automation and Information Systems,\\
Technical University Munich, Germany\\
	\email{vogel-heuser@tum.de}
}

\authorrunning{Alenazi et al.}

\maketitle

\context
An automated production system (aPS) is a design-to-order, custom-built system (e.g., a wood-working plant) that is operated over decades and undergoing continuous changes in its mechatronic and software parts~\citealenazi{BVH-JSS15}. Traceability has long been recognized as key to change management: Identify who needed the change and why, locate where the change should be applied, reason about the change impact, etc. However, the change management support, and for that matter, the traceability itself, has been exclusively scoped \emph{within} the software engineering artifacts and processes~\citealenazi{JCH-FOSE14}.


\questions
The challenge facing aPS evolution is to \emph{go beyond} the software (engineering) boundary because the changes commonly involve both the software (e.g., source code) and the mechatronics (e.g., sensors and actuators). Moreover, the changes of the software and those of the mechatronics are constantly out of sync. For example, replacing a malfunctioned sensor may not trigger any software change, but replacing it with a better one may lead to software updates so that the new sensor's data could be read and processed more accurately.

A root cause of the challenge is that aPS evolution requires different specialities to be communicated and coordinated. At least three types of expertise are needed: mechanical engineering, electrical engineering, and software engineering. Traceability in aPS evolution, then, is truly a systems phenomenon and no longer scoped only within software development. While creating and maintaining traceability is important, too much traceability can hurt the autonomy of the aPS stakeholders (mechanical engineers, electrical engineers, and software engineers). Thus, not only do we need a new form of \emph{systems traceability}, but we shall also address such questions as when, who, to whom, and how much associated with systems traceability in the context of aPS evolution.

\ideas
We envision the systems traceability serves as both an interface and a protocol between software and mechatronic parts of aPS. Our idea is to base Protos~\citealenazi{Chopra-RE14} to create a representation of aPS traceability.

Protos models the requirements of a socio-technical system as commitments between interacting parties rather than the goals of each individual. A socio-technical system involves interactions between autonomous principals facilitated by technical artifacts, including software. Emphasizing interactions, rather than machine/software specifications, has shown to promote openness and accountability in the era of the IoT~\citealenazi{Chopra-WWW16}.

Although Protos can handle a single set of changing requirements, its support for continuous aPS changes is insufficient. We therefore propose to augment Protos with specific traceability links with the objective to enhance aPS sustainability. In our vision, the traceability links will not only substantiate the commitment and its refinement, but also enable the identification of conflicts undermining the new needs' fulfillment, the reasoning of requirements-induced technical debt, and the creation of innovative requirements for the aPS. Our ongoing work with a bench-scale aPS, namely the PPU (Pick \& Place Unit)~\citealenazi{BVH-TR14}, shows some preliminary results and insights.

\contrib
 We contribute a formal yet purposeful framework based on Protos for defining systems traceability and using it to sustain aPS changes. The framework contains the formal notations to establish commitments between aPS social entities, the refinements of commitments towards fulfilling the changes, and the traceability as argumentation to guide how the refinements should be carried out.

\future
 Building on and extending our work, we envision the next ten years will shift traceability from inside the software's boundary to the mechatronics that software invariably interacts with and that co-evolves with software.

\bibliographystylealenazi{modifiedplain}
\bibliographyalenazi{abstracts/bibs/alenazi}

\newpage
\title{Semantically Enhanced Software Traceability\\Using Deep Learning Techniques}

\author{Jin Guo \and Jinghui Cheng \and Jane Cleland-Huang}

\institute{Department of Computer Science and Engineering\\
	University of Notre Dame, South Bend, IN, USA\\
	\email{\{jguo3, JinghuiCheng, JaneClelandHuang\}@nd.edu}
}

\authorrunning{Guo et al.}
\titlerunning{Semantically Enhanced Software Traceability}

\maketitle

\context
Total automation of high-quality trace creation and maintenance has been identified as one of the important research task in the Grand Challenges of Traceability report \citeguo{RERoadmap}. To achieve this goal, researchers have devoted significant amount of effort in the past decades -- more than half of the recent traceability-related research papers have focused on the automated solutions for trace link creation and/or maintenance \citeguo{DBLP:conf/icse/Cleland-HuangGHMZ14}. 

\questions
Those solutions have included information retrieval approaches, machine learning, heuristic based techniques, and AI swarming algorithms. However, the results have been mixed, especially when applied to industrial-sized datasets, where acceptable recall levels above 90\% can often only be achieved at extremely low levels of precision. One of the primary obstacles that current automated approaches fail to overcome is the semantic gap between pairs of related artifacts. To address this problem, more intelligent tracing solutions incorporating domain-specific concepts, and reason intelligently about relationships between artifacts are vitally needed.

\ideas
The goal of our current work is to utilize deep learning to develop a scalable, portable, and fully automated solution for bridging the semantic gap that currently impedes the success of trace link creation algorithms. Our solution is designed to automate the capture of domain knowledge and the artifacts' textual semantics with the explicit goal of improving accuracy of the trace link generation task. The proposed approach includes two primary phases.  First, we learn a set of \textit{`word embeddings'} for the domain using an unsupervised learning approach trained over a large set of domain documents. The approach generates high dimensional word vectors that capture distributional semantics and co-occurrence statistics for each word. Second, we use an existing training set of validated trace links from the domain to train a Tracing Network to predict the likelihood of a trace link existing between two software artifacts. Within the Tracing Network, we adopt a \textit{Recurrent Neural Network (RNN)} architecture to learn the representation of artifact semantics. For each artifact (i.e. each regulation, requirement, or source code file etc.), each word is replaced by its associated vector representation learned in the word embedding training phase and then sequentially fed into the RNN. The final output of RNN is a vector that represents the semantic information of the artifact. The Tracing Network then compares the semantic vectors of two artifacts and returns the probability that they are linked. The Tracing Network has proven useful to effectively construct semantic associations between artifacts, and to deliver significantly higher Mean Average Precision (MAP) scores than state-of-art methods when evaluated on one large industrial dataset. It furthermore notably increased both precision and recall.

\contrib
Given an initial training set of trace links, our Tracing Network is fully automated and highly scalable. We have demonstrated that deep learning techniques can be effectively applied to the tracing process. We see this as a non-trivial advance in our goal of automating the creation of accurate trace links in industrial-strength datasets.

\future
Despite the advances contributed by our current work, there are still many open challenges that need to be be addressed in order to fully leverage semantic analysis for tracing purposes.  Deep learning techniques require large amounts of data currently not available in the public domain.  Current community datasets are too small for training deep learning models.  Nevertheless, there are several ways ahead such as working to acquire large datasets of trace links or exploring the use of unsupervised techniques instead of supervised ones. Another challenge is the need for exploring ways to reuse existing models in new projects. Differences in project characteristics require models to be retrained in new contexts, thus requiring ways to facilitate the transfer of models across projects with minimal retraining.  Finally, deep learning techniques are not a silver bullet. It may be difficult to design and train a deep learning model that is sophisticated enough to capture all the nuances of complex artifacts in the software engineering domain. We should therefore explore hybrid solutions that leverage heuristic approaches in conjunction with deep learning. Despite these challenges we believe that focusing on semantic-based solutions provides promise for future advancements in the field.

\bibliographystyleguo{modifiedplain}
\bibliographyguo{abstracts/bibs/guo}

\newpage
\title{Evolving Requirements to Source Code Trace Links in Safety-Critical Domain}

\author{Mona Rahimi \and Jane Cleland-Huang}
	
	\institute{Department of Computer Science and Engineering\\
			University of Notre Dame, South Bend, IN, USA\\
	\email{mrahimi@nd.edu; JaneClelandHuang@nd.edu}}

\authorrunning{Rahimi et al.}
\titlerunning{Evolving Requirements to Source Code Trace Links}

\maketitle

\context
In safety-critical domains, certification guidelines prescribe that trace links must be created between software artifacts such as hazards, faults, requirements, code, and test cases in order to demonstrate that all hazards have been fully addressed and mitigated in the deployed system~\citerahimi{DBLP:journals/infsof/NejatiSFBC12,DBLP:conf/icse/RempelICSE14,CSER,FDA2}. Software-intensive systems constructed in safety-critical domains therefore involve extensive traceability efforts. While trace links are problematic and time-consuming to create~\citerahimi{Olly1995,DBLP:journals/tse/RameshJ01}, they can be even more challenging to maintain accurately as the system evolves~\citerahimi{journals/cacm/DomgesP98}.

\questions
The inevitable fact is that quality of existing trace links in a software-intensive system can dramatically degrade over time as the system evolves. Changes that frequently occur in core artifact types such as source code, architectural models, and requirements can render previously valid trace links invalid. Further, an additional set of trace links may be required to keep trace information up to date. Constantly maintaining trace links as software-intensive systems evolve over time is cumbersome, error-prone and costly. Moreover, outdated trace links invalidate safety-cases, which rely on trace links and associated data to provide evidence for system safety. 

\ideas
To address the aforementioned \emph{trace link degradation} problem, we emphasize the need for effective trace link maintenance. Our solution focuses on evolving trace links according to changes occurred in the system.  

We implemented and evaluated a Trace Link Evolver (TLE) which is designed to automatically evolve trace links between unstructured natural language requirements and source code. The process starts with an initial version containing requirements, source code, and trace links. When a new version of requirements and source code is introduced, TLE detects predefined change scenarios that have occurred between the two versions. Each change scenario is associated with a corresponding set of trace link heuristics responsible for uncovering  links that need to be created and/or deleted in order to evolve trace links. 

We evaluated TLE through conducting a series of three experiments with requirements and java source code. The first evaluation included a controlled experiment with eleven java developers making modifications to two java applications.  Secondly, we evaluated several variants for implementing TLE in a live dronology project environment including integrating analyst feedback and the use of class- versus method-level links. Finally, to evaluate TLE’s ability to evolve trace links in a larger industrial context, we applied it across 27 releases of the Apache Cassandra Database Management System. Results show that the trace links evolved using TLE are significantly more accurate compared to those generated solely using information retrieval techniques.

\contrib
 TLE's heuristics are designed to achieve extremely high recall without unduly sacrificing precision. We demonstrated that the process of evolving trace links is less expensive and more accurate than generating links from scratch. We also showed the benefits of including human decision making processes within the TLE evolution cycle. Our final experiment showed that TLE is capable of scaling up to large-sized projects.

\future
The problem of trace link evolution represents a pressing, and challenging area where further research is required. Our results clearly demonstrate the benefits gained when evolving links as part of the natural maintenance process. In future work we will conduct further evaluations including additional, diverse systems, with hundreds of releases. Furthermore, we aim at exploring the evolution of trace links across other types of safety artifacts in addition to requirements and source code.

\bibliographystylerahimi{modifiedplain}
\bibliographyrahimi{abstracts/bibs/rahimi}

\newpage
\title{How Eye Tracking Benefits Software Traceability}
\author{Bonita Sharif}

\institute{Department of Computer Science and Information Systems \\ 
Youngstown State University\\ 
Youngstown, Ohio USA 44555\\
\email{bsharif@ysu.edu}
}

\authorrunning{Sharif}

\maketitle

\context
The paper presents the role of eye tracking in software traceability link creation and evolution. Two studies are presented to illustrate the feasibility of the approach. The main traceability challenges our work addresses are related to a) making traceability ubiquitous and b) making traceability scalable. \\
When traceability is ubiquitous, it becomes almost effortless. Our premise is that traceability can be captured while developers are working on their daily tasks. An eye tracker can unobtrusively collect eye gazes on artifacts while a developer works. At the end of a work session, the system can record links that were derived based on what artifacts were looked at. \\
The use of an eye tracker to capture and maintain traceability links can be done at any level of granularity.  It can also be done among many different types of artifacts, all of which are accessible within the user's work environment.  This makes the approach scalable. 

\questions
The main research questions we have addressed are:

\begin{itemize}
	\item RQ1: Is eye gaze a viable and feasible source to recover traceability links?
	\item RQ2: How do the results of gaze-based algorithm compare to the results of information retrieval methods?
	\item RQ3: How do we determine the ground truth when generating traceability links via eye gaze?
\end{itemize}

We believe using eye tracking is very different from traditional methods of how traceability is currently done. The eye-tracking enabled software traceability environment we developed namely, iTrace (http://seresl.csis.ysu.edu/iTrace), can collect eye gaze data necessary to generate a software traceability model and generate candidate traceability links among a wide variety of artifacts and granularity levels. This entire process is conducted without interrupting the developer's workflow.

\ideas
In 2014, we conducted a pilot study to determine if eye tracking was a feasible approach (RQ1) to derive traceability links  \citesharif{Walters:2014}. The system we used was iTrust, where we injected one-line bugs into the system so developers would perform bug localization tasks while they their gaze was recorded. The links were compared to a set of true links created before the study. The links generated by our gaze algorithm were then ranked by another set of developers. The results were promising in terms of recall (60\% on average) however precision needed improvement (30\% on average). 
In 2016, we presented a comparative study \citesharif{Sharif2016} to address RQ1 and RQ2. The comparison was between links generated from IR methods and from eye gazes. The eye gaze approach outperformed standard LSI and VSM approaches and reports a 55\% precision and 67\% recall on average for all tasks when compared to how the developers actually fixed the bug (i.e., commits). The gaze algorithm is crucial to pruning out irrelevant and stray glances. 

One of the main concerns related to this type of approach is how to evaluate ground truth (RQ3).  Commits are typically used as a source to compare results from IR based methods. Eye tracking data provides insights that go beyond entities that were changed (i.e. reflected in commits). To establish eye gaze as ground truth, more work needs to be done and the approach needs further validation in the traceability community.

\contrib
The main premise of using eye tracking towards helping build and maintain software traceability is that is makes traceability part of the work being done – thus making it somewhat effortless, ubiquitous, and scalable (with tool support).

\future
The gaze-link algorithm can be further refined to learn based on each developer's behavior and expertise. It should be able to consolidate links based on a common link repository. Another important aspect is to increase the trustworthiness of the links generated, directly related to the trusted challenge and determining ground truth.  
Finally, we plan on incorporating our method into work environments such as Eclipse and generate potential candidate links after each developer session.

\bibliographystylesharif{modifiedplain}
\bibliographysharif{abstracts/bibs/sharif}


\newpage

\part{Trace Link Usage}
\author{Session Chairs:\\Markus Borg and Bonita Sharif}
\title{\sessionheader{Session 3:\\ Trace Link Usage}}
\institute{}
\authorrunning{Session 3}
\titlerunning{Session 3}
\maketitle
\setcounter{section}{0}
\vspace{40pt}





In this session, the focus was on the usage of traceability links. Creating and storing links is great but not enough. Developers do not necessarily use links simply because they exist. Traceability links are not always trusted to support their developers in their work. Furthermore, Cleland-Huang mentioned that only 30\% of traceability papers are in the usage field. Clearly, this needs to change in the future as the usage of traceability tools become more prevalent. 

During the opening of the session, Markus Borg presented results from an empirical study \citesessionThreeIntro{borgicpc17} on software engineers' information seeking behavior in change impact analysis, including `tracing' as an explicit seeking strategy. Based on 14 interviews with engineers in the process automation domain, i.e., in a development context in which traceability is carefully maintained to comply with safety standards, the authors conclude that all engineers do not consider trace links particularly useful. The recommendation is that the development organization should attempt to make the links more accessible -- to ensure traceability return on investment.  

The session continued with three talks on traceability usage. Each talk was fifteen minutes long. We give a brief summary of each talk next. Goodrum et al.'s work deals with exploring a visual interface for traceability data to help the user. They engage developers in interviews and observations. The end goal is to propose novel interactive ways to present and consume trace data. Farrar et al. had a unique application of traceability techniques.  It dealt with using traceability for policy checking in networks and in particular the NetSecOps project.
Kokaly et al. discussed trace links usage in the automotive industry. The paper discusses how safety cases should be linked to evidence in the system that could come in the form of testing, analyses, and model checking to name a few. 

After the three talks, we organized an open activity followed by a discussion. This exercise lasted approximately an hour. The open activity involved posing five questions to the audience. The questions were placed on paper in the four corners of the room (question 4 and 5 were put on the same sheet of paper). The audience was asked to address the questions by placing their idea to a solution for the question on a sticky note near the question. At the end of the session, one person was chosen to give a summary of the post-it ideas/solutions for each of the five questions. 

The five questions along with the responses received from the audience were as follows:

\begin{enumerate}
    \item For which software engineering activities do practitioners use trace links?
        \begin{enumerate}
        \item Regression testing 
        \item Maintenance
        \item Test coverage
        \item Requirements coverage
        \item Quality assessment
        \item Bug localization
        \item Continuous integration (CI) environments
        \item Bug fixes
        \item Feature Location
        \item Program Understanding
        \item Verification and Validation
        \item Onboarding project - learning new system
        \item Generating Test Oracle
        \item Requirements Reuse
        \item Test Refactoring
        \item Low-level Design
        \end{enumerate}
    \item How do we make links actionable and useful in some way? Define use cases.
        \begin{enumerate}
        \item Better visualizations
        \item High precision
        \item Integration with IDE and part of workflow
        \item Embed during tasks
        \item Task-specific traces/filtering
        \item Trace queries
        \item Tools that are designed for use in context
        \item TODO list generation
        \item Easy to understand
        \item Well maintained
        \item Better HCI
        \item Automated Evolution
        \item Lightweight links and traceability
        \end{enumerate}
    \item What do practitioners think about using trace links? Utility? Cost/benefit ratio?
        \begin{enumerate}
        \item Indifferent smiley
        \item Uninteresting?
        \item Cost is high, Benefit is low in the short term, high in the long term
        \item Very important but on high cost
        \item Only useful when they cause no additional effort
        \item Too time consuming
        \item Too expensive
        \item Costly too maintain
        \item Not worth the effort
        \item How I can use it?
        \item Some like it, some don't
        \item Cost is way greater than benefit
        \item Unclear responsibility
        \item Just for Validation, Certification, Compliance
        \end{enumerate}
    \item How and when are trace links updated?
        \begin{enumerate}
        \item Updated manually when changes occur.
        \item In practice: Before certification by an unlucky guy. More ideally: Automatically as changes occur.
        \item After the fact between versions
        \item At runtime while making changes
        \item Rarely
        \item When a commit occurs (evolve)
        \item When forced/external pressure
        \item Ad-hoc/Delayed
        \end{enumerate}
    \item How and where are trace links stored?
        \begin{enumerate}
        \item Bug tracking systems
        \item Database + XLS + DOC + EMF
        \item In many different places: Jira (Code to requirements), Inside test cases (references), Spreadsheets
        \item Within assets (internal strategy)
        \item XML, CSV, repositories
        \item DOORS
        \item Excel
        \item Jira
        \item Lightweight database
        \end{enumerate}
\end{enumerate}


Finally at the end of the session, we recap on where we are at and how well we were doing. We wanted to know if we are making progress on turning traceability into a useful and appreciated concept. We were interested in finding road blocks to adoption of traceability. We talked about what our next goals should be. Hopefully this brainstorming session, where people looked at answers to the above questions, helped spark a new research interest in trace link usage. The more usable we make our traces, the better industry will be willing to adopt the process and practice of using them.

\bibliographystylesessionThreeIntro{modifiedplain}
\bibliographysessionThreeIntro{abstracts/bibs/session3_intro}

\newpage

\title{Improving Usability of Safety Critical \\Requirements Traceability}

\author{Micayla Goodrum \and Ronald Metoyer \and Jane Cleland-Huang}

\institute{Department of Computer Science and Engineering\\
University of Notre Dame\\
Notre Dame, IN\\
\email{\{Mgoodrum, rmetoyer, janeclelandhuang\}@nd.edu}
}

\maketitle

\context
Despite the benefits traceability can bring to a software project, it can be difficult to achieve in practice \citegoodrum{DBLP:journals/software/MaderJZC13} due to the cost and effort required to create and maintain trace links, and the lack of effective tools and techniques for visualizing and leveraging trace data in ways that support software engineering tasks \citegoodrum{RameshJ01}. Maalej surveyed existing methods for representing trace data and found that in practice, the most common way to enter and show traceability information was the trace matrix , which results in usability issues as the system becomes larger~\citegoodrum{DBLP:conf/refsq/LiM12,Cleland-HuangGHMZ14}. 

\questions
The goal of this current research is to characterize existing software trace data and developer needs and to propose novel interactive representations to make trace data more usable for safety analysts and software developers. 

\ideas
We performed an informal study in which we asked developers to use traceability data to identify the impact of a new requirement request.  The participants were given a series of Change Requests for a small safety-critical system, we informally observed which artifacts our users leveraged to perform the task.  We also found that users commonly used the artifacts to look up information and identify connections.  For example it was common for participants to make comments such as ``...I am looking at the trace links to see if there are any hazards already linked to increasing the amount of drones...''   These domain-specific tasks are used in the next section to identify a subset of abstract tasks.

From the informal study, we map the domain-specific safety tasks and data to a common set of tasks and data, for which there may already be well-known visual representations to support them. We identified the data types from the artifacts that users were given in the observation, where we found that many were either Nominal or textual.

From analysis of user quotes, such as those presented in the previous section, we identified two main tasks:  retrieve artifacts and connect/relate artifacts. Using the example quote in the previous section, we can determine that the user is attempting to \emph{retrieve or look up} a particular hazard and \emph{connect or relate} that hazard to the source code.  These abstract tasks, \emph{retrieve} and \emph{connect}  are used in the next level, encoding, to decide how to visually represents the artifact data.

The next level of the nested model requires that we choose appropriate encodings for the data to be represented and tasks to be carried out.  

Users must be able to \emph{retrieve} safety artifacts and identify the \emph{connections} between them.  The spatial channel is the most effective encoding for both quantitative as well as nominal data, thus we use position to encode the artifact ids themselves (e.g. requirements, source, design decisions, etc.).  According to Gestalt principles \citegoodrum{peterson13} one of the most effective means for visually grouping two elements together is through the use of connecting lines.  Therefore a path can be encoded by connecting source-target pairs from the tracematrix. The path that is encoded using our system, is a forward and backwards trace from the FEMCA that is associated with the requirements connected to the code of interest. The resulting representation, a node-link diagram, is no surprise and fully justified by our design decisions. We present a node-link layout with the nodes placed horizontally, left to right.  The forward trace path is always displayed above the nodes.  Upon mouseover, the second pass links would display below the nodes.

\contrib
Information visualization may prove valuable for supporting developers making safety critical software updates.  We present the application of a principled design decision model to a small safety critical software system. We also proposed encoding principles that can be applied to better visualize safety critical systems. 

\bibliographystylegoodrum{modifiedplain}
\bibliographygoodrum{abstracts/bibs/goodrum}

\newpage

\title{NetSecOps and Policy Checking}
\subtitle{An Application of Traceability Techniques}
\author{David Farrar \and Jane Huffman Hayes \and\\ Gabrielle Adkins\and James Griffioen \and Cody Bumgardner}
\authorrunning{Farrar et al.}

\institute{Computer Science Department\\
	University of Kentucky\\
	Lexington, Kentucky, USA\\
	\email{\{hayes, griff\}@cs.uky.edu,\\  \{cody, david.farrar, gabrielle.adkins\}@uky.edu}
}

\maketitle

 \context Campus networks often support substantial levels of open access that are needed for faculty, staff, students, and guests to collaborate and carry out their work.  Unlike networks which are carefully locked down and monitored (such as corporate or medical networks), campus networks can be particularly difficult to secure, and difficult to know when the network has been compromised and by whom.  One way to address this is to define human-readable network access and use policies that describe the University’s acceptable use policies and reflect the desire of the University to protect sensitive data and the network itself while allowing open access to the network as much as possible.  These policies then need to be translated into the technical specifications that drive the network’s routers and switches.  Recently, software defined networking (SDN) techniques make it possible for these high level policies to be decomposed into very specific lower level SDN rules installed in SDN-enabled switches  Moreover, the actual network flows can then be logged.  It is vital to check the resulting logs to ensure that high level organizational policies and programs/rules have indeed been followed.  
 
\questions
 Automated traceability techniques can be used to assist with checking of natural language text policies against SDN applications (such as those that make use of the OpenFlow protocol) as represented in the National Science Foundation (NSF)-funded NetSecOps project.
 
\ideas We are applying components developed for TraceLab \citefarrar{cleland2012toward}, a traceability experimentation framework developed under an NSF grant, to perform checking between network flow logs and higher level security policies.  We are using these pre-processing components:  splitting, stemming, and stop word removal.  We are using these processing components:  build corpus and generate trace links.  We are developing our own components for importing source and target artifacts (some artifacts are represented as json).  We are developing our own components to perform heuristic-driven checks of the logs against the policies.  To date, we are finding the challenges of the project to be related to the: lack of high level and intermediate level policies, learning curve as we work in a new domain, as well as the continuing evolution of the program that will generate logs (and thus the continuing evolution of the format of the artifacts that need to be traced/checked). 

\contribfuture
 We plan to capitalize on already existing traceability techniques and components in Tracelab and develop new components as needed.  We are gaining domain knowledge from our network security partners at the University of Kentucky and the University of Utah, while hopefully imparting software and requirements engineering knowledge to them.  In addition to applying traceability techniques to a new domain, we are gaining knowledge about how to interact with domain experts as we do so.

\bibliographystylefarrar{modifiedplain}
\bibliographyfarrar{abstracts/bibs/farrar.bib}


\newpage
\title{Trace Links and Their Use in Automotive Software Safety Assessment}

\author{Sahar Kokaly}
\authorrunning{Kokaly}

\institute{
    General Motors, Canada\\
	McMaster Centre for Software Certification, Canada\\
	University of Toronto, Canada\\
    \email{sahar.kokaly@gm.com | kokalys@mcmaster.ca |  skokaly@cs.toronto.edu}
}

\maketitle

\context
The pervasiveness of software in all aspects of human activity has created special concerns regarding issues such as safety, security and privacy. Governments and standard organizations (e.g., ISO) have responded to this by creating regulations and standards that software must comply with. In the automotive domain, ISO 26262 is the de facto standard for assessing functional safety of road vehicles. As part of demonstrating compliance, an artifact, called a \emph{Safety Case}, which shows that the system indeed satisfies the property imposed by the standard (in this case safety), is produced. The recommended safety development process in ISO 26262 is to perform a Hazard Analysis and Risk Assessment, where potential hazards and residual risks are identified, and use them to identify \emph{Safety Goals}. The latter are then refined into \emph{Functional Safety Requirements}, which are in turn refined to \emph{Technical Safety Requirements} that refer to particular software and hardware components in the system, and are decomposed into \emph{Software Safety Requirements} and \emph{Hardware Safety Requirements}, respectively. The point is to be able to ultimately produce an argument (the \emph{Safety Case}) which demonstrates that each of the safety goals has been met, by eventually linking them to \emph{evidence} in the system. This evidence can come in the form of test results, analyses, model checking results, expert opinion, etc.

\questions
Like other systems, automotive software systems naturally evolve due to a variety of reasons including adding, removing or modifying
features, fixing bugs, or improving quality. In this context,
their safety cases also need to co-evolve. A necessary step in this coevolution is performing an impact
assessment to identify how changes in the system
affect the safety case. The question we are interested in is the following: Given a safety case for an original system, can we aid the safety engineer in constructing a safety case for the evolved system by reusing the components of the original safety case as much and as soundly as possible, thus reducing the overall revision cost incurred by the safety engineer?

\ideas We introduced a model-based impact
assessment approach, that given a model of the system (often described
as a \emph{megamodel} -- a model with many heterogenous models and
relationships between them), a safety case, a
traceability relationship between the system megamodel and the safety
case, a delta representing a change in the system, and appropriate
slicers for each of the model types in the system megamodel, the
approach is able to produce an annotated safety case that reflects the
impact of the system changes on the safety case elements.  We recently
identified a set of techniques for improving precision of impact
assessment; many of these techniques focus on improving the
system-safety case, standard-system and standard-safety case
traceability.

\contrib Three techniques directly relate to traceability. (1): Increasing the granularity of traceability between the system and the safety case. This allows us to identify finer-grained impact of system changes on the safety case elements.(2): Identifying sensitivity of safety case to system changes. By attaching additional knowledge to the trace links that identifies under which cases a change in the system should actually impact a linked element in the safety case, and when such a change can be ignored (e.g., name changes), we are able to reduce the cost of revising the safety case by eliminating unnecessary revision.
(3): Understanding standard-system and standard-safety case traceability. ISO 26262 includes additional information about how ASILs (Automotive Software Integrity Levels), usually assigned to safety case goals, are related to ISO 26262 Work Products, which refer to system models used as evidence to support the safety case. Knowing this traceability significantly enhances the impact assessment and produces more precises results.
We are actively working on extending our model management framework MMINT (https://github.com/adisandro/MMINT) to include safety cases and model management operators for them (e.g., safety case slice) as well as explicit trace links between the safety case and the standard/system.

\future We believe that our approach can be used for impact assessment in general, and not just for safety case co-evolution. One application for this is in design space exploration, to enable answering what-if questions about the impact of changes on safety cases. In this context, we aim to study the effect of changes, other than just system changes, on the safety case and understand what types or trace links are required to support them. An interesting challenge we currently face is when the system change involves adding elements
which we cannot handle since we are unable to automatically
discover links to the safety case.

\ack This work is done in collaboration with Rick Salay, Marsha Chechik (University of Toronto) and Mark Lawford, Tom Maibaum (McMaster University).



\newpage

\part{Real-World Applications of Traceability}
\author{Session Chairs:\\Sahar Kokaly and Michael Vierhauser}
\title{\sessionheader{Session 4:\\ Real-World Applications of Traceability}}
\institute{}
\authorrunning{Session 4}
\titlerunning{Session 4}
\maketitle
\setcounter{section}{0}
\vspace{40pt}
\titlerunning{Real-World Applications of Traceability}


The transformation from concepts, ideas, and prototypes developed in research to approaches and tools that can be used in practice is often a long and cumbersome road.
Many approaches that have been proven valuable in theory are perceived as too theoretical lacking real-world applicability \citesessionFourIntro{huangAcademia,wohlinAcademia}.
Two of the 'grand challenges of traceability'  are to (i) make traceability \emph{ubiquitous}, i.e. always there when needed as an integral part of a company's processes and tools and (ii) make traceability \emph{valued}, i.e., raising the awareness of traceability and its value in real-world applications.

In this session -- \emph{Real-World Applications of Traceability} -- we focused on the practical application of trace planning, trace usage, trace evolution, and trace maintenance approaches. This session included three very diverse abstracts where the authors report on their experience and ideas on applying traceability in real-world projects.

First, Markus Borg presented his initial results on applying traceability in the domain of autonomous driving vehicles. Dealing with safety-critical systems, in their work they employ machine  learning techniques and neural networks in the context of  creating trace links. 
Then, Michael Vierhauser reported on his work on runtime monitoring in large-scale system of systems. His work aims at  uncovering and establishing trace links between runtime and design-time artifacts which can provide  valuable support for diagnosing violations of requirements at runtime. Finally, Patrick M{\"a}der presented results of an experiment on how traceability is perceived in practice in real-world development projects.

 \bigskip
To foster awareness of how existing traceability approaches developed by researchers can be used in practice and can be valuable for industry we performed a 3MT-like presentation\footnote{\url{https://threeminutethesis.uq.edu.au}} contest as part of our session. Participants were given 3 minutes each to present their work and describe how and why their approach can be of practical value.
Mona Rahimi presented \emph{Trace Link Evolver (TLE)}\citesessionFourIntro{Rahimi_TLE} 
an approach for supporting trace link and their evolution across multiple versions of a software system.
Second, Jan-Philipp Stegh{\"o}fer presented Eclipse Capra\footnote{\url{https://eclipse.org/capra}}, a  traceability management tool for the Eclipse development environment.  It provides capabilities for defining traceability information models, creating  traceability links between various  artifacts  as well as visualization capabilities. Next, Bonita Sharif presented an approach for  creating and maintaining  trace links using eye-tracking. Sugandha Malviya presented an approach for collecting and moreover analyzing requirements related queries to software development projects, and finally, Tingting Yu presented an approach supporting software testing.

The results and approaches presented in this session and the resulting discussions provided the basis for the subsequent in-depth discussion in our Industry Transfer breakout session (cf. Breakout Groups).

\bibliographystylesessionFourIntro{modifiedplain}
\bibliographysessionFourIntro{abstracts/bibs/session4_intro}


\newpage
\title{Traceability and Deep Learning –- Safety-critical Systems with Traces Ending in Deep Neural Networks}
\author{Markus Borg\inst{1} \and Cristofer Englund\inst{2} \and Boris Dur\'an\inst{2}}

\institute{RISE SICS AB\\
	Lund, Sweden\\
	\email{markus.borg@ri.se}
	\and
	RISE Viktoria AB\\
	Gothenburg, Sweden\\
	\email{cristofer.englund|boris.duran@ri.se}
}

\authorrunning{Borg et al.}
\titlerunning{Traceability and Deep Learning}

\maketitle

\context
Deep learning has enabled significant advances in various applications in the recent decade. Thanks to increased computational power in combination with abundant data, a steady stream of deep learning success stories has been published, particularly in computer vision and natural language processing. In the automotive industry, deep learning is considered the most promising solution to realize the image recognition and perception required to allow autonomous driving (AD). However, an autonomous car is a safety-critical system that must comply with the ISO 26262 standard, thus the car manufacturers must show that all safety requirements are satisfied in the delivered product –- a typical traceability use case.

\questions
Developing safety-critical systems with an a deep learning component bring new challenges, no matter whether the system is trained once on a batch of training data before deployment, or if the system allows online learning. As opposed to traditional software components implemented manually in source code, a typical deep learning model is opaque, i.e., it cannot be subjected to code reviews or tested exhaustively. A deep neural network is a complex structure, often represented by hundreds of millions of weighted parameters that distribute information. Thus, it is currently not possible for a human to interpret how and where decisions are made. If the implementation of a safety requirement for AD, e.g., stopping for crossing pedestrians, is traced to a deep learning image recognition component – how do you motivate that the system is reasonably safe and thus ready for safety certification? Do the trace links need to penetrate the deep learning component, i.e., point at specific parts of the neural network? Or should the development organization trace to specific training examples that have been used to train the network, such as an image set of one million crossing pedestrians in various traffic environments? There is a strong need for traceability research in the automotive industry, aligned with the goal of ``purposed traceability’’, i.e., traceability fit for the purpose of certifying safety-critical systems with ML component.

\ideas
We plan to conduct a literature review on V\&V for autonomous systems, potentially safety-critical, implemented using machine learning components. At the same time, we are charting industry practitioners’ perception of challenges involved in developing systems that rely on deep learning. We believe that the closest area of related work is component-based software engineering (CBSE) for safety-critical systems, introducing safety concepts such as safety cages and redundancy. We aim to identify how findings in CBSE could be adapted to develop safety-critical systems with deep learning components. 

\contrib
We will use snowballing to conduct a systematic literature review of state-of-the-art V\&V for safety-critical systems with machine learning components. Concurrently, we will organize a series of focus groups with industry practitioners to survey the corresponding needs from the automotive industry when developing vehicles with AD features. Finally, we will combine our understanding of state-of-the-art and state-of-practice to identify promising solutions and needs for future research.

\future
Considerable software engineering research effort is needed to enable cost-efficient development of safety-critical systems that rely on deep learning. We are confident that traceability will remain the foundation for the safety case argumentation, and our future work will explore ``purposed traceability’’ in the context of developing AD. More explicitly, our goal is to help industry understand where the traces from the safety requirements should end to allow safety certification. Three different alternatives all open avenues for future work, i.e., traces ending in 1) a ``safety cage’’ around an opaque deep learning component, 2) inside a human-interpretable model of a deep neural network, and 3) in training examples used to train and test the deep learning model.

\ack
This work was carried out within the SMILE project financed by Vinnova, FFI, Fordonsstrategisk forskning och innovation under the grant number: 2016-04255.

\newpage

\title{Establishing Trace-Links for Runtime Diagnosis Support in System of Systems}

\authorrunning{Vierhauser et al.}
\titlerunning{Establishing Trace-Links for Runtime Diagnosis}

\author{Michael Vierhauser\inst{2} \and Rick Rabiser\inst{1} \and Paul Gr\"unbacher\inst{1} \and Jane Cleland-Huang\inst{2}}

\institute{Christian Doppler Laboratory MEVSS, ISSE\\
	Johannes Kepler University Linz, Austria\\
	\email{\{rick.rabiser | paul.gruenbacher\}@jku.at}
	\and
	Department of Computer Science and Engineering\\
	University of Notre Dame, South Bend, IN, USA\\
	\email{\{mvierhau | janeclelandhuang\}@nd.edu}
}

\maketitle

\context
Many industrial software systems today are systems of systems~(SoS) characterized by decentralized control; support for multiple platforms; inherently volatile and conflicting requirements; continuous evolution and deployment; as well as heterogeneous, inconsistent, and changing elements \citevierhauser{Maier98,Nielsen2015SoS}. Such SoS cannot fully be tested during development time: interactions between the SoS and its environment can only be checked during operation when all of its software systems, including legacy and third-party software, and the hardware interoperate for the first time.
 Techniques such as requirements monitoring thus have to be used to observe such systems at runtime to detect deviations from their requirements.\\
\questions However, the focus of existing monitoring approaches is mainly on detecting violations of expected behavior, while support for subsequent diagnosis of violations is rather limited and often even neglected \citevierhauser{VierhauserSLR}. Diagnosis is particularly challenging in SoS, which are characterized by complex heterogeneous architectures and a slew of different development and testing tools.\\
\ideas
With our work we aim to improve support for monitoring and diagnosing complex software-intensive systems and their requirements. Specifically, we want to complement requirements monitoring with traceability, to both, design-time and runtime artifacts. This includes algorithms and methods, supported by tools, and combining existing state-of-the-art runtime monitoring techniques with runtime traceability support for different types of artifacts. \\
\contrib
  Our work targets (i) uncovering relevant artifacts for different types of systems, stakeholders, and purposes as well as activities that can be performed on these artifacts to support diagnosis of violated requirements; (ii) support for establishing trace links between the artifacts and runtime information; (iii) support for maintaining artifacts and their trace links when the monitored systems and their requirements evolve; and (iv) tools for engineers, project managers, or maintenance personnel to efficiently use the provided knowledge to detect, analyze and fix requirements violations at runtime.
 Our current work focuses on a traceability-supported framework for diagnosing requirements violations in large-scale, heterogeneous systems based on an existing requirements monitoring model (RMM) \citevierhauser{VierhauserRE2015} complemented by a system artifact model (SAM). We use the SAM for collecting artifacts and linking them to specific activities that can be performed by arbitrary tools (e.g., opening a source code artifact in an external IDE) and to establish trace links between elements of the RMM (such as requirements and constraints) and different kinds of artifacts (such as source code, configuration files, and requirements documents).
 Such trace links can support engineers and maintenance staff in analyzing  violations of requirements at runtime by providing additional information available in the linked artifacts.\\
 \future As part of our ongoing research we plan to broaden our scope and support different types of artifacts and diagnosis activities and furthermore add automation support for automatically uncovering trace links to important artifacts.

\bibliographystylevierhauser{modifiedplain}
\bibliographyvierhauser{abstracts/bibs/vierhauser}

\newpage
\title{Benefits and Challenges of Software Traceability in Development Projects}

\author{Patrick M\"ader}
\authorrunning{M\"ader et al.}

\institute{Technische Universit\"at Ilmenau\\Software Engineering for Critical Systems Group\\Ilmenau, Germany\\
\email{patrick.maeder@tu-ilmenau.de}
}

\maketitle

 \context
 Software traceability is a required component of many software development processes. Advocates of software traceability cite advantages like easier program comprehension and support for software maintenance (i.e., software change). However, despite its growing popularity, for a long time there existed no published evaluation about the usefulness of requirements traceability. It is important, if not crucial, to investigate whether the use of requirements traceability can significantly support development tasks to eventually justify its costs.

\questions
We thus conducted a controlled experiment with 71 subjects re-performing real implementation tasks on two third-party development projects: half of the tasks with and the other half without traceability \citemaeder{pma1,pma2}.  Our findings show that subjects with traceability performed on average 24\% faster on a given task and created on average 50\% more correct solutions -- suggesting that traceability not only saves effort but can profoundly improve software implementation quality. For a follow-up study, we selected 24 medium to large-scale open-source projects and focused especially on the discovered effect for implementation quality \citemaeder{pma3}. We quantified for each developed component of each software project, the degree to which a studied development activity was enabled by existing traceability and set this metric in relation to the number of defects that occurred in a component.

\ideas 
We found that traceability significantly affects the defect rate in a component.

\contrib
Overall, our results provide for the first time empirical evidence that traceability significantly improves implementation speed as well as implementation quality during software development.

\future
Further studies are required to understand the benefits and challenges of traceability in support of other development activities, e.g., impact analysis.

\ack
We are funded by the German Ministry of Education and Research (BMBF) grants: 01IS14026A, 01IS16003B, by DFG grant: MA 5030/3-1, and by the EU EFRE/Th{\"u}ringer Aufbaubank (TAB) grant: 2015FE9033.
\bibliographystylemaeder{modifiedplain}
\bibliographymaeder{abstracts/bibs/maeder}




\newpage
\part{Traceability Datasets and Benchmarks}
\author{Session Chairs:\\Mona Rahimi and Carlos Bernal-C\'ardenas}
\title{\sessionheader{Session 5:\\ Traceability Datasets and Benchmarks}}
\institute{}
\authorrunning{Session 5}
\titlerunning{Session 5}
\maketitle
\setcounter{section}{0}
\vspace{40pt}

\titlerunning{Traceability Datasets and Benchmarks}


Lack of publicly available data sets is one inhibiting factor to do research in the area of treaceability and more specifically in the area of trace link maintenance. To assess and compare the performance of evolutionary algorithms, an initial set of trace links is required, as well as a reference set of evolved trace links for evaluation purposes. Finding such datasets on public domain is one of the challenges that researchers in this area encounter.

A benchmark for traceability provides support for evaluation and validation of trace link creation and maintenance techniques. In addition, benchmarking traceability facilitates the comparison of different traceability techniques and assist research community to reach consensus over various approaches~\citesessionFiveIntro{chen2013development,ben2011towards}. Building benchmarks for evaluating traceability methods and techniques, is stated as one of the grand challenges in traceability~\citesessionFiveIntro{dekhtyar2007benchmarks}.

In this session, we specifically focused on addressing these two challenges in the area of traceability. First, the challenge of publicly available datasets and second, the challenge of benchmarking traceability. We raised potential discussion topics such as what are the requirements for a benchmark for traceability tasks? What are the essential properties that evaluation methods should possess? Do we need to include human factor in traceability benchmark? What are the most serious restrictions on building a traceability benchmark? and finally what are the benefits of benchmarking traceability.

Carlos Bernal-C\'ardenas first presented a demo of TraceLab as a potential benchmark for traceability\footnote{\url{https://github.com/CoEST/TraceLab}}. Later, Jane Cleland-Huang familiarized the participants with available datasets on CoEST website\footnote{\url{http://sarec.nd.edu/coest/datasets.html}}. During this session, we performed a panel with four panelists, Dr. Giulio Antoniol from École Polytechnique de Montréal, Dr. Jane Hayes from University of Kentucky, Dr. Nan Niu from University of Cincinnati, and Dr. Patrick M{\"a}der from Technische Universit{\"a}t Ilmenau. During the panel, our panelists shared their thoughts and point of views on where we are currently standing with respect to traceability benchmarking and datasets. While all panelists agreed with existing challenges in finding publicly available datasets, some discussed that benchmarking traceability has its own challenges and restrictions. Finally, to wrap up the session, we divided participants into breakout groups to further discuss the concerns mentioned by our panelists. The detailed outcome of breakout groups is reported in our Datasets and Benchmarking breakout session (cf. Breakout Sessions).


\bibliographystylesessionFiveIntro{modifiedplain}
\bibliographysessionFiveIntro{abstracts/bibs/session5_intro}






\part{Additional Contributions}


\newpage
\clearpage
\thispagestyle{empty}
\pagebreak
\hspace{0pt}
\vfill

\begin{center}
\begin{minipage}{.90\textwidth}
{\centering\fontfamily{qtm}\fontsize{31}{15} \selectfont Additional Contributions}
\end{minipage}
\end{center}

\vfill
\hspace{0pt}
\pagebreak
\title{Enabling Domain-specific Traceability\\ with Eclipse Capra}

\author{Salome Maro \and Jan-Philipp Stegh\"ofer}

\institute{Chalmers $|$ University of Gothenburg\\\
\email{salome.maro@gu.se, jan-philipp.steghofer@gu.se}
}

\authorrunning{Maro et al.}
\titlerunning{Enabling Domain-specific Traceability\\ with Eclipse Capra}

\maketitle

\context
In the development of complex systems, traceability provides benefits such as facilitating change impact analysis, program comprehension, tracking of project progress and artifact reuse~\citesteghoefer{bouillon2013survey}. 
In order to leverage these benefits, a development organization needs to plan for how traceability should be established. This includes eliciting the goals of the stakeholders and providing traceability solutions that achieve these goals. The goals of traceability vary from organization to organization and sometimes from project to project. Eliciting these goals and devising context-specific solutions is not a trivial task and requires investment by the development organization. This is because, there are no ready-made goals and solutions available from research that practitioners can use to bring traceability into their organizations. Therefore, when planning for traceability, the organization needs to elicit the goals, define a traceability information model based on the goals and the artifacts in the organization and acquire and customize tools based on the goals and information model. Moreover, it is not clear how the introduction of traceability activities (e.g., creating, maintaining and using traceability) integrates into the existing development processes of the development organization.

\ideas
To accommodate these varying needs of stakeholders, we have developed the traceability management tool Eclipse Capra\footnote{\url{https://eclipse.org/capra}}~\citesteghoefer{maro2016capra}. It provides the ability to define traceability information models, create traceability links to development artifacts (including requirements, design documents, tests, and code), as well as visualization of these links.

As the requirements of different development organizations and even different development projects vary significantly, Eclipse Capra is both configurable (to allow users to select functionalities that they need) and extensible (to allow addition of new artifacts that can be traced and addition of new traceability features). However, the time and effort required to elicit traceability requirements from stakeholders and configure the tool for usage in a real life environment can be high. To reduce this time, we aim to provide reasonable defaults for different domains that can serve as a starting point for customization. We also aim to study how traceability supported by Eclipse Capra can be introduced and integrated in ongoing industrial development processes. 

We aim to develop traceability information models that are relevant for different domains and contexts, investigate how the information models can be instantiated in  different development organizations, and how this impacts the existing development processes. 
To achieve this, we will conduct case studies with development organizations where we will elicit their needs, define the traceability information model, use the tool to instantiate the information model and  allow the users to create and use traceability links.  We will conduct interviews and observations of traceability activities (creation, maintenance, use, exchange etc.) in order to understand how they integrate with the development processes.  

\contrib
The contribution of our research will be twofold: a set of traceability information models mapped to domain-specific/context-specific goals and recommendations for how traceability activities should be integrated with various development processes. We will therefore be contributing to both the purposed and configurable challenge from \citesteghoefer{gotel2012quest}. The set of information models can be used by development organizations who want to establish traceability as a starting point when defining their information models. The knowledge gathered from integrating traceability in real life projects will also give insights on how traceability activities fit into different development processes and act as a source of best practice advise for practitioners. 

\future
We are currently planning the first integration of Eclipse Capra into a development process with a company in the financial domain. Other initiatives are underway to investigate traceability information models for multi- and many-core development and for product line engineering.

\bibliographystylesteghoefer{modifiedplain}
\bibliographysteghoefer{abstracts/bibs/steghoefer}

\newpage

\title{An Information Theoretic Approach for Traceability Link Retrieval}

\author{Saket Khatiwada \and Miroslav Tushev \and Anas Mahmoud}
\authorrunning{Khatiwada et al.}

\institute{Division of Computer Science and Engineering\\
Louisiana State University\\
Baton Rouge, LA, 70803\\
	\email{\{skhati1, mtushe1, amahmo4\}@lsu.edu}
}

\maketitle

\context
 As software systems grow in size and complexity, traceability can become a tedious and time-consuming process. To minimize the manual effort, contemporary traceability tools utilize Information Retrieval (IR) methods for automated support. IR methods exploit the textual content of software artifacts to automatically capture and rank potential trace links. In this work, we propose a new paradigm of information-theoretic IR methods to support various traceability tasks in software systems. These methods, including Pointwise Mutual Information (PMI) \citemahmout{m1} and Normalized Google Distance (NGD) \citemahmout{m2}, exploit the co-occurrence patterns of words in software systems to reveal hidden textual semantic dimensions that other methods often fail to capture \citemahmout{m3}. Our objective is to establish more accurate semantic similarity relations (potential traceability) between software artifacts.
 
\method
Six datasets from different application domains are used to conduct our analysis. The proposed methods are compared against classical IR methods that are often used in software traceability research, including: Vector Space Model (VSM), as a representative of lexical matching methods, Latent Semantic Indexing (LSI) \citemahmout{m4}, as a representative of semantically enabled IR methods, and Jensen-Shannon Model (JSM) \citemahmout{m5}, as a representative of probabilistic IR methods.

\results
The results show that information-theoretic IR methods fail to outperform the classical IR methods in smaller datasets. However, they are able to outperformance other, more computationally-expensive, semantic IR methods in datasets with larger number of traceability links.

\contrib
This research adds to the existing body of research on IR-based traceability by investigating new paradigm of light-weight IR methods that have the potential to scale up to larger and more complex software systems.
\newpage
\future
Our experimental datasets will be enhanced with more software systems sampled from a broad range of application domains. Our objective is to evaluate the performance of our proposed methods over various configuration settings. Furthermore, a set of working prototypes that implement our findings will be developed. Working prototypes will allow us to conduct long term usability studies to gain a better understanding of our methods’ scalability, usability, and score of applicability.

\bibliographystylemahmout{modifiedplain}
\bibliographymahmout{abstracts/bibs/mahmout}

\newpage

\title{Feature-Oriented Traceability}
\author{Thorsten Berger}
\institute{Chalmers $|$ University of Gothenburg\\
\email{thorsten.berger@chalmers.se}}

\authorrunning{Berger}

\maketitle

\context
Features are commonly used to describe the functional and non-functional aspects of a system. Features are abstractions over implementation assets and understood by many different roles, including domain experts, architects, and developers. As such, features are often used for communication, planning, and keeping an overview understanding of a system.
Some software-engineering methods advocate the explicit use of features, such as feature-driven development (FDD) and software product line engineering with feature modeling. Especially the latter requires abstractions (\textit{features}) to cope with complex product lines---portfolios of system variants tailored towards specific requirements, such as different market segments, hardware, or non-functional properties (e.g., performance or energy consumption).

Variants are typically developed using clone\&own\,\citeberger{berger.ea:2013:survey}---that is, copying and adapting existing variants to new requirements. This strategy is simple and allows experimenting with new ideas and rapidly prototyping variants. However, it does not scale well, and maintaining variants quickly becomes costly. Then, variants often need to be migrated to an integrated product-line platform. Such a platform is often configurable and allows deriving variants by selecting dedicated features in a configurator tool. Unfortunately, the product-line migration is costly and risky, requiring architectural and organizational changes, as well as recovering features and their locations.

We believe that recording features and their locations early during clone\&own facilitates feature maintenance and evolution, including a later platform migration.
Established feature traceability would avoid the expensive recovery of feature traces (e.g., when modifying, removing or reusing features) or allow analyzes and predictions on the level of features.

\questions Our past and current work targets three questions.

Q1: \textit{What are effective feature-traceability methods?} Most established methods retroactively recover feature traces and are heavyweight, imposing significant setup (e.g., creating a traceability database) and ongoing costs (e.g., updating traces in the database). We believe that effective methods should encourage developers to continuously record features during development and should be lightweight, facilitating easy integration into engineering processes with low overhead.

Q2: \textit{What are the benefits and costs of feature traceability?}
The cost of creating and maintaining traceability links should be lower than the expected benefit. We conjecture that feature traceability does not only enhance feature-oriented engineering activities and analyzes, but also support traceability tasks not related to features, such as change-impact analysis of requirements. We believe that some of the low-level and fine-grained traceability links (e.g., between requirements and test cases) can be replaced by fewer, but higher-level traceability links based on features as pivotal (and intuitive) elements.

Q3: \textit{How to leverage expert knowledge?} Most feature-traceability techniques focus on the fully automated recovery of feature locations, typically using informa\-tion-retrieval techniques. Yet, these techniques are inaccurate and have not found widespread adoption. Since features are highly domain-specific\,\citeberger{berger.ea:2015:feature}, expert (developer) knowledge should be taken into account. We believe that effective feature-traceability methods are hybrid, supporting developers with partial automation.

\ideas
Towards Q1, we conceived a lightweight feature-traceability approach relying on embedded annotations. During implementation, developers record features and the feature locations as annotations in code\,\citeberger{ji2015annotations}. Our feature dashboard tool allows exploiting annotations by extracting and visualizing them\,\citeberger{andam2017florida}. Towards Q2, we studied costs and benefits of embedded annotations\,\citeberger{ji2015annotations}, which showed that their cost is low compared to their benefit for feature maintenance and evolution. We currently investigate the benefit of feature traceability for other traceability tasks, such as change-impact analysis for requirements. We strive to conduct user studies on feature traceability compared to traditional traceability, investigating how the notion of features is perceived and under what conditions feature traceability is beneficial. Towards Q3, we currently conceive a hybrid approach that learns from past annotations and recommends new ones. Developers should proactively record annotations, but should be reminded about potentially forgotten traces (e.g., upon commit). We systematically experiment with machine-learning techniques by replaying the history of feature annotations in a case study.

\contrib
 We contribute to the grand challenges of traceability by raising the abstraction level of traceability to the (intuitive) notion of features, by arguing for feature traceability as a way to provide more short-term benefits of traceability in order to improve its general acceptance, and by providing empirical data on the costs and benefits of feature traceability.

\future
While our work is motivated by the needs of variant engineering in clone-based development, we believe that making features explicit and establishing feature traceability also enhances single-system engineering. Investigating traceability strategies around the notion of features in single systems is a valuable future research direction.

\bibliographystyleberger{modifiedplain}
\bibliographyberger{abstracts/bibs/berger}

\newpage

\title{Datasets in Software Traceability Research}
\subtitle{Current State and Characteristics}

\author{Waleed Zogaan \and Palak Sharma \and Mehdi Mirahkorli}

\institute{Software Engineering Department\\
    Rochester Institute of Technology, USA\\
    \email{\{waz7355,ps2671,mxmvse\}@rit.edu}
}

\authorrunning{Zogaan et al.}
\maketitle

\context
Datasets are crucial to advance automated software traceability research. Acquiring such datasets come in a high cost and require expert knowledge to manually collect and validate them. Obtaining such software development datasets has been one of the most frequently reported barrier for researchers in the software engineering domain in general~\citezoogan{Liebchen:2008:DSD,shepperd2013data}. This problem is even more acute in the area of requirements traceability which is crucial in safety critical and highly regulated application domains~\citezoogan{DBLP:conf/icse/CHuangGHMZ14}. Therefore, the main motivation behind this work is to analyze the current state of art of datasets used in the field of software traceability. We aim to identify the challenges related to datasets and try to mitigate them by first gaining some insight throughout analyzing the characteristics of currently used datasets by the researchers.

\questions
Following a systematic literature review (SLR), we identified 200 research papers focused to acquire knowledge about the currently used datasets in the field of software traceability. Utilizing our gathered knowledge, we aim to answer these questions: a) What characteristics makes one dataset used more frequently than the others? Does there exist such a rationale for dataset selection? b) Does quality of a dataset play a role in choosing a dataset as a case study? If so, how do we define quality?

\ideas
To identify and collect the datasets used in software traceability community, we conducted a systematic literature review (SLR) of all published full papers with \textit{empirical} and \textit{automated} software traceability \textit{theme}. We followed the guidelines that were established by Kitcheman et al.~\citezoogan{kitchenham2007guidelines} for SLR in Software Engineering. As per our data exploration so far, we found that a) Dataset in safety critical domains - Aerospace and Healthcare - are more frequently used than others. b) The size of the datasets differently enormously for Open Source, Industrial and Academic projects. c) Most datasets are used to study traces between code to Requirements and other non-code artifacts. d) Overall 40\% of datasets are not available for reuse, mainly belonging to industrial and academic categories. Our work aims to identify the challenges related to datasets by analyzing their characteristics and trying to create a benchmark for assessing their quality.

\contrib
Throughout this study, we aim to provide new insight on the characteristics of datasets used in our community, present a new quality assessment means to reason about traceability datasets, and reveals tacit information about a large number of datasets used in the community which can highlight the path for addressing threats to validity of the research conducted in this area. 

\future
We plan to expand our analysis and cover different aspects related to datasets quality in order to come up with a framework that enable researchers to assess their datasets quality.

\bibliographystylezoogan{modifiedplain}
\bibliographyzoogan{abstracts/bibs/zoogan}

\newpage

\title{A Natural Language Interface for Trace Queries}

\authorrunning{Lin et al.}

\author{Jinfeng Lin \and Jane Cleland-Huang}

\institute{
	Department of Computer Science and Engineering\\
	University of Notre Dame, South Bend, IN, USA\\
	\email{\{jlin6@nd.edu | janeclelandhuang\}@nd.edu}
}

\maketitle

\context
Trace links, which are created in many projects -- especially safety-critical ones, are often underutilized in practice, due to the fact that  stakeholders often lack knowledge in how to leverage them. However, if leveraged effectively, trace links can be used to construct cross-artifact queries in support of diverse software engineering activities such as impact analysis, test regression selection, and coverage analysis \citelin{DBLP:journals/sosym/MaderC13}.  

\questions
One of the major adoption barriers, that impedes the efficient use of trace links in practice, is the lack of accessibility to underlying trace data \citelin{DBLP:conf/re/GotelCHZEGA12}. In order to retrieve data and to compose it in meaningful ways, stakeholders have to construct and execute queries that cut across potentially heterogeneous, distributed, and interdependent data sources. Unfortunately, many project stakeholders lack the skills to perform this task. To alleviate this problem, we propose a Natural Language (NL) Interface, TiQi~\citelin{DBLP:journals/re/PruskiLGRC15}, capable of interpreting and executing trace queries in a software project environment. TiQi, enable users to verbalize or write queries using their own words, thereby removing one of the barriers for using existing trace links and querying software projects.

\ideas
TiQi accepts natural language queries and transforms them through a series of steps into an executable SQL query.  To support this multi-phase transition, the TiQi solution is constructed using a pipe-and-filter architecture in which each filter is dedicated to a certain aspect of the NL translation problem.  A NL query entering the pipeline will be processed and refined until an executable SQL query can be generated by the final filter. 
Individual filters are designed to support tasks such as \textit{Query Intention Analysis}, \textit{Keyword Matching}, and finally \textit{SQL generation}. 
For query intention analysis, we parse the syntax and context of the natural language query to determine the type of the query. For example, based on the grammar we can easily distinguish queries which enumerate data entries from those which count data entries. For keyword matching,  words and phrases in the initial query are mapped to specific artifacts, attributes, or data values. To achieve high matching accuracy, an extensive and domain specified vocabulary is indispensable.
\newpage
\contrib Our research focus is delivering an efficient and accurate NL query interface in order to aid project stakeholders in issuing queries. By reducing the overhead of accessing and using existing trace data, project stakeholders will be able to leverage existing traceability data in order to ask meaningful analytic questions about the project data.

\future
As part of our ongoing and future work we work on continuously improving TIQI and its constituent components. First of all, we aim at supporting more complicated queries, such as queries with sub-clauses. 
Converting sub-clauses into sub-queries enables users to express their queries more eloquently. Secondly, we are currently working on refining of the vocabulary TIQI  can rely on, to significantly reduce the keyword mismatching rate.

\bibliographystylelin{modifiedplain}
\bibliographylin{abstracts/bibs/lin}

\newpage

\title{Building Ontology \\to Support Trace Query Terms}

\authorrunning{Liu et al.}

\author{Yalin Liu \and Jane Cleland-Huang}

\institute{
	Department of Computer Science and Engineering\\
	University of Notre Dame, South Bend, IN, USA\\
	\email{\{yliu26@nd.edu| janeclelandhuang\}@nd.edu}
}

\maketitle

\context
Software and systems engineering projects produce diverse data artifacts such as requirements, design, source code, test cases, and project plans. If queried effectively, this data can be leveraged to deliver project intelligence. However, stakeholders find it challenging to construct formal queries in order to retrieve and leverage this data. Although natural language (NL) interfaces can alleviate this problem, they need the ability to process diverse, technical terminologies embedded in user queries and then to discover associations between different, yet related, terms. Recent research has shown that ontology can serve as an internal mediator for identifying relationships between query terms \citeliu{DBLP:conf/icse/LiC13,guo2016tackling}. For purposes of querying software projects, there are two distinct domains -- first, the domain of the project itself (e.g. communications or healthcare), and second, the software engineering domain which introduces its own terminology (e.g. terms such as use-case, specification, and clone). Ontology can be useful in both domains; however, here we focus on software engineering terms.

\questions
Imagine a software project in which a user asks ``Which requirements have failed their acceptance tests?'' Given a traceability information model (TIM) that includes an artifact type labeled `requirements' and another labeled `acceptance tests', any NL interface should be able to successfully map the query to the appropriate artifacts. On the other hand, if the user referred to requirements as `use-cases' or `user stories', the NL interface could only perform the mapping if it were aware that use-cases and user-stories were alternate ways of describing requirements. This type of information can be documented in an ontology. However, a general ontology, such as WordNet, does not include many commonly used Software Engineering terms. The question we address here is whether we can construct an ontology of software engineering terms that can be used to improve the ability of a NL interface to understand user queries.


\ideas
The first step in building a domain-specific ontology involves collecting large amounts of software engineering documents from which we can extract commonly used terms and phrases. We build our domain-specific corpus upon this data. Then, we leverage  the lexical, contextual, and syntactic information to discover associations between various phrases. We then store this data in an ontology and use it to map NL queries to terms in the TIM. In turn, the improved mapping enables more accurate trace queries.

\contrib
The contribution of this work will be an extensive ontology of Software Engineering terms and their associations.  This ontology will support accurate transformation of NL queries into executable trace queries.

\future
After we implement our domain-specific ontology, we will embed it into natural language interfaces, such as TiQi~\citeliu{DBLP:conf/re/PruskiLAOARC14} and evaluate its efficacy for improving the accuracy at which NL queries can be successfully interpreted and executed.
\bibliographystyleliu{modifiedplain}
\bibliographyliu{abstracts/bibs/liu}



\newpage
\part{Discussion \& Breakout Groups}
\newpage
\setcounter{section}{0}

\pagestyle{fancy}
\renewcommand{\headrulewidth}{0pt}
\lhead{}
\rhead{}
{\fontfamily{qtm}\fontsize{22}{15}\selectfont 
{\noindent{Discussion \& Breakout Groups}}
}
\\
\\

\renewcommand{\thesection}{\oldthesection)}

\section{Industry Transfer}
\label{breakout_industry}
\authorrunning{Grand Challenges of Traceability 2017}
\titlerunning{Industry Transfer}


As discussed in our session on real-world applications of traceability, there are still quite a few problems and obstacles that impede the use of traceability approaches and the adoption of research prototypes in industry. The goal of this breakout session thus was to uncover, and furthermore, discuss these existing obstacles and propose ideas an concrete action items addressing these issues.
The discussion in the different focus groups resulted in three concrete areas we deem as important for future traceability research in the context of technology transfer: the perceived \emph{value} of traceability in industry and the lack of perceived \emph{importance} of traceability itself as part of the development process; the lack of existing \emph{standards} regarding artifacts and formats; and the current problems when \emph{publishing results} from industrial collaborations.

\subsection*{Make Traceability Valuable \& Make Traceability Sexier}
Despite the wide variety of traceability approaches and the advances over the last 10 years, industry still seems to not be fully convinced of the the value of traceability in their daily (software development) business. Apart from domains where traceability is prescribed by legal requirements and standards \citebreakoutIndustry{is1992software,iso26262,medevice} approaches have not, and still are not widely used in practice.
However, not only the perceived lack of value of traceability approaches needs to be addressed but also the topic itself seems to lack a certain 'sexiness' factor 
We thus think that the following action items can provide the foundation for increasing the acceptance of traceability in industry and facilitate the use and adaption of traceability research approaches in industry:
\newline
First of all, companies need to be convinced, that traceability in fact adds additional value to their development processes and ultimately better quality in their products. It is therefore important to 
 (i) \emph{emphasize the return on investment (ROI)} of traceability and to highlight benefits when introducing traceability in their company. 
This goes hand in hand with (ii)  \emph{introducing and promoting success stories} where traceability has been successfully applied in industrial projects. These success stories should highlight the benefits of traceability and report on both, short term and long term advantages.
This could, for example, include conducting cost-benefit studies to emphasize the benefits of traceability and furthermore subsequent large-scale and longitudinal studies investigating the benefits over a longer period of time. From our point of view, the key challenge is to find and present scenarios where traceability results in short term benefits, making it more attractive to use approaches in industry.
Furthermore, to increase awareness (both in industry and in research) of the traceability research area, the  we think that (iii) traceability needs a \emph{re-branding and a focus on more recent technologies}. This includes for example social media presence, announcing conferences, workshops, and newly published papers and approaches and the involvement of participants from industry, e.g., at conferences, workshops and symposiums.   

\subsection*{Invest in Publishing Results}
In order to convince industrial users and promote industrial use-cases, research should also tackle this issue. Besides reporting and publishing articles using (and evaluating approaches with) small research data sets, researchers should aim to incorporate real-world systems and data sets.
However, the problem with these real-word systems is that oftentimes certain restrictions regarding the publication of results arise. Companies are not willing (or not able) to grant rights to researchers to publish company-internal data sets or even results and findings based on this data. Since the publication of research results and their public discussion is vital for academic research, this apparent clash  of interest between industry and academic research needs to be addressed. 
We therefore propose to invest in \emph{obfuscation techniques and community standards} that allow in reporting on results without disclosing sensitive information on the one hand, but still provide enough information to be valuable and enable the validation of results. Once prepared and published, this data (e.g., in form of a data set) should be made publicly available, allowing a broader group of researchers to experiment with the data and validate research results. By collecting and systematically providing such real-world data sets to the research community we think that this initiative could encourage other researchers to prepare and share their data sets within the traceability community.

\subsection*{Invest in Standards}
Lastly, as we have observed, the heterogeneity of artifacts and formats involved in any traceability maintenance setting impedes the use of approaches in different application areas or domains. Although there might not be a "one size fits all" solution, we believe that investing in \emph{standards} that regulate which formats are used for what artifacts in which domains, and guidelines for unification of formats between application areas, will drastically improve the efficiency and applicability of many traceability techniques that have been proposed in the literature. 


\bibliographystylebreakoutIndustry{modifiedplain}
\bibliographybreakoutIndustry{abstracts/bibs/breakout_industry}




%
%

\newpage
\section{Datasets and Benchmarking}
\label{breakout_dataset}




During the discussion, we highlighted three main problems facing us as a community in traceability benchmarking and datasets. This group focused on problems oriented to datasets that involve \emph{obtaining data, data verification and data feedback}.

\subsection*{Obtaining Data}
This relates to the problems that researchers face with benchmarks since there is a dearth of artifact (e.g., test cases, bug reports) diversity as well as domain specific datasets (heterogeneity) to generalize the results. Most datasets only provide links from source to requirements. Another problem related to obtaining datasets is that some datasets are proprietary which may mean that there needs to be some possible solutions to anonymize datasets. Help might also be needed in packaaging of the dataset. 

\subsection*{Data Verification} 
This problem relates to the lack of domain experts that can validate traceability links or have participated in the creation of the ground truth datasets. It is important for researchers to know whether the benchmark was reviewed either by experts, senior developers, or students.  Therefore, each dataset should include metadata describing the nature of the review and data. 

\subsection*{Data Feedback} 
This relates not only on the use of datasets but also reporting back whether links work for specific traceability tasks including a replication package. Moreover, the research community benefits from knowing what has been tried before and replication packages that can run to compare new approaches with baselines.

\section*{Actionable items}
As a community, it is important to come up with actionable items to consider for each of the above areas in order to move the field forward.  The actionable items that will eventually lead to a potential solution are listed below. 

\begin{itemize}
    \item \textbf{Obtaining Data}: One possible action item in the near future should involve creation of incentives to report back on datasets. One such option is to provide recognition to papers with a special dataset or artifact badge both on the paper and on the website.   In the call, specific instructions should be given on how to package and anonymize the data. The COEST leadership team could also host a good citizen annual award for the best dataset. Furthermore, include not only positive results but negative results detailing the tasks in which the links did not work well. Last, but not the least organize a special issue about tracing test cases to requirements for datasets including replication packages. Finally, sessions should be held on determining ways in which datasets can be curated and anonymized. These actionable items can help in adding domain diversity, improve data set anonymity and packaging. 
    
    \item \textbf{Data Verification}: A possible solution to verifying data sets would be to run an open source study at various different points of interest with varying levels of population samples. Replication is key to verifying links and maintaining quality. It also increases our confidence in the links. A grant proposal specifically targeting data verification might be a possible solution to increase verified datasets. 

    \item \textbf{Data Feedback}: One way to improve data feedback is to provide incentives to researchers to use existing data in order to report back on the results they received. Data feedback also helps in maintaining the quality of the links and in supporting further use of them. 
\end{itemize}

\section*{Summary}
In summary, work needs to be done in all three areas of obtaining data, verifying data, and data feedback. Some issues such as quality of the dataset crosscuts all three areas and some actionable items overlap one or more areas. For example, running an open source study on different artifacts across many sites helps both obtaining data and verifying the data across the different sites.  A community effort and willingnes to share datasets is essential to solving some of these problems.

%
%
%
%
%
%
%
%
%
%

\newpage


\end{document}